\documentclass[journal,draftclsnofoot,onecolumn,12pt,twoside]{IEEEtran}
\usepackage{amsmath,amsfonts}
\usepackage{array}
\usepackage{comment}
\usepackage[caption=false,font=normalsize,labelfont=sf,textfont=sf]{subfig}
\usepackage{textcomp}
\usepackage{stfloats}
\usepackage{url}
\usepackage{verbatim}
\usepackage{graphicx}
\usepackage{cite}
\usepackage{soul}
\usepackage{booktabs}
\usepackage{hyperref}

\usepackage{url}
\usepackage{ifthen}
\usepackage{cite}
\usepackage{braket}
\usepackage{amssymb}
\usepackage{amsthm}
\usepackage{graphicx}
\usepackage{dsfont}
\usepackage{tikz}
\usepackage{algorithm}
\usepackage{algpseudocode}
\usepackage{comment}

\newtheorem{theorem}{Theorem}
\newtheorem{definition}{Definition}
\newtheorem{proposition}{Proposition}
\newtheorem{lemma}{Lemma}
\newtheorem{remark}{Remark}

\newcommand{\R}{\mathcal{R}}
\newcommand{\F}{\mathbb{F}_2}
\newcommand{\wH}{\mathrm{w}_{\mathrm{H}}}
\newcommand{\Ann}{\mathrm{Ann}}
\newcommand{\supp}{\mathrm{supp}}

\begin{document}

\title{Weight-$6$ Generalized Bicycle Codes with Growing Dimension and  Minimum Distance}

\author{%
  \IEEEauthorblockN{Thamadathil Aswanth, Sakshi Dang, Gyanendra K. Verma, and Navin Kashyap}\\
  \IEEEauthorblockA{Dept. Electrical Communication Engineering, Indian Institute of Science\\
                    {\small aswantht@iisc.ac.in, sakshidang10@gmail.com, gkvermaiitdmaths@gmail.com, nkashyap@iisc.ac.in }  } 
}



   \maketitle

\begin{abstract}
Generalized Bicycle (GB) codes form an important class of quantum error-correcting codes constructed within the Calderbank-Shor-Steane (CSS) framework using block-circulant matrices. In this work, we study the polynomial description of GB codes to obtain explicit characterizations of their stabilizers and logical operators, and derive a formula for the minimum distance in terms of polynomial weights. Using this framework, we prove a distance lower bound for a broader construction and prove equality in two parameter regimes. We present explicit constructions of infinite families of GB codes with weight-$6$ stabilizer generators, arising from an asymmetric $(2,4)$ split of row weights in the two circulant blocks. These constructions preserve the low-density parity-check (LDPC) property and achieve dimension and minimum distance both growing as $\Theta(n^{1/3})$, where $n$ denotes the blocklength. 
\end{abstract}


\section{Introduction}\label{sec:intro}

Quantum error correction is a central ingredient of scalable fault-tolerant
quantum computation~\cite{shor1995scheme}. Quantum low-density parity-check
(qLDPC) codes are particularly attractive because they admit stabilizer
generators of bounded weight, with each physical qubit participating in a
bounded number of checks~\cite{breuckmann2021qldpc}. Although asymptotically
good qLDPC code families are well known~\cite{panteleev2022asymptotically,leverrier2022quantum},
constructions with short blocklengths and small check weights remain important
for near-term practical implementations. Bivariate Bicycle (BB) codes~\cite{bravyi2024high} and Generalized Bicycle (GB) codes~\cite{Pryadko,Panteleev2021degeneratequantum} are two prominent classes of codes that exhibit good dimension and minimum distance for short blocklengths. GB codes are a structured class of CSS codes defined by two commuting circulant
matrices. Their sparse check
patterns and cyclic structure have motivated finite-length studies and
implementation proposals on neutral-atom arrays~\cite{viszlai2023matching}.

GB codes are instances of the two-block group-algebra codes over a
cyclic group~\cite{2bga}. Their circulant blocks can be represented by
polynomials in $\mathbb{F}_2[z]/\langle z^N-1\rangle$, where $N$ is the
circulant block size, and the corresponding quantum blocklength is $n=2N$. This representation
supports algebraic descriptions of stabilizers and commuting Pauli
operators. Davenport \emph{et al.} developed a cyclic-submodule framework
for GB codes and studied structured logical-operator
bases~\cite{davenport2026generalized}.

Determining the minimum distance remains a difficult step in the design
of GB and BB codes. The minimum-distance problem is NP-hard for general
stabilizer codes~\cite{kapshikar2023hardness}, and therefore numerical searches play an
important role in identifying useful finite-length examples. Randomized
information-set methods provide distance upper
bounds~\cite{pryadko2022qdistrnd}, whereas exact certification requires a
matching lower bound, obtained, for example, by integer-programming
computation or exhaustive distance-verification
procedure~\cite{wang2026coprime,rowshan2026logical}.
At the asymptotic level, locality-based bounds rule out asymptotic goodness
for GB code families with fixed check weight~\cite{GBCodeDistance_symmetry,arnault2025variant}.
For explicit families, exact distances have been determined for
weight-$4$ constructions~\cite{GBCodeDistance_symmetry,arnault2025classification}.
Dastbasteh \emph{et al.} presented an infinite family of $[[2(s + 1)(2s + 1), 2, 2(s + 1)]]$ GB codes with weight-$6$ stabilizers, for odd integers $s\geq 3$ \cite{dastbasteh2025gb}.
These results motivate constructions in which the number of encoded qubits
and the minimum distance both increase.

In this work, we study weight-$6$ GB codes with an unequal allocation of check weight
between the two blocks of physical qubits. Each $X$-type check acts on two
qubits in the first block and four in the second, with these roles reversed
for the $Z$-type checks. We refer to this allocation as a $(2,4)$ split. Within this class, we impose an
algebraic relation between the two check blocks that fixes the number of
encoded qubits while retaining freedom in the choice of check patterns.
This permits distance analysis at prescribed dimension and fixed check
weight. Our main contribution is an explicit
family with a distance lower bound that is attained for infinitely many
parameter choices.

\paragraph*{Contributions}
\begin{enumerate}
\item \emph{An explicit family with exact distance:}
We construct a family of weight-$6$ GB codes, establish a lower bound on
its minimum distance, and identify two subfamilies for which this
bound is attained (Theorem~\ref{thm:2-4}). In particular, for every integer
$t\geq 2$, the construction includes a code with parameters
\[
  [[\,2(t+1)t^2,\,2(t+1),\,t\,]],
\]
so both the number of encoded qubits and the minimum distance grow as
$\Theta(n^{1/3})$, where $n$ denotes the blocklength. Table~\ref{tab:one-parameter-families} in Section~\ref{subsec:one-parameter} gives the parameter choices for several specializations of Theorem~\ref{thm:2-4}. All of these subfamilies have connected Tanner graphs.
The distance analysis bounds the weights of nontrivial logical operators while
accounting for stabilizer degeneracy.

\item \emph{An algebraic description for distance analysis:}
Building on polynomial parametrizations of GB logical
operators~\cite[Eq.~(12)]{GBCodeDistance_symmetry} and the submodule
framework~\cite{davenport2026generalized}, we provide a self-contained
description adapted to our construction. We characterize the logical-$X$
operators and identify those belonging to the stabilizer group
(Proposition~\ref{prop:logical_modules} and Lemma~\ref{lem:logXmeetStabX}). The resulting expression for the minimum distance is then used to establish the bounds for our explicit family.
\end{enumerate}

The parameters $[[2(t+1)t^2,2(t+1),t]]$ coincide with those of $t+1$
independent square toric codes. However, Proposition~\ref{prop:indecomposable} shows that the constructed code admits no tensor-product decomposition across a nontrivial partition of its physical qubits.

The remainder of the paper is organized as follows.
Section~\ref{sec:prelims} introduces the notation and recalls CSS and GB
codes. Section~\ref{sec:main} develops the polynomial description of
stabilizers, logical-$X$ operators, and minimum distance.
Section~\ref{sec:explicit} proves Theorem~\ref{thm:2-4}, characterizes
Tanner-graph connectedness, and presents explicit one-parameter families.

\section{Notations and Preliminaries} \label{sec:prelims}
Let $\mathbb{F}_2$ denote the binary field, $\mathbb{Z}_+$ the set of positive integers, and $\mathbb{C}$ the field of complex numbers. For a positive integer $m$, $\mathbb Z_m$ denotes the groupof integers modulo $m$ under addition.

For binary vectors $\mathbf{a}=[a_0, a_1, \ldots, a_{n-1}]$ and $\mathbf{b}=[b_0, b_1, \ldots, b_{n-1}] \in \mathbb{F}_2^n$, their dot product is defined as $\mathbf{a}\cdot\mathbf{b} = \sum_{i=0}^{n-1} a_i b_i$. The Hamming weight of a vector $\mathbf{a} \in \mathbb F_2^n$ is given by $\mathrm{w}_H(\mathbf{a}) := \#\{i \mid a_i \neq 0\}$. For a binary linear code $\mathcal{C} \subseteq \mathbb{F}_2^n$, its dual code is defined as $$
\mathcal{C}^\perp := \{\mathbf{c}^\prime \in \mathbb{F}_2^n \mid \mathbf{c}\cdot \mathbf{c}^\prime=0 \text{ for all } \mathbf{c} \in \mathcal{C}\}.
$$

Let 
\[
    X := \begin{bmatrix} 0 & 1 \\ 1 & 0 \end{bmatrix}, \quad 
    Y := \begin{bmatrix} 0 & -i \\ i & 0 \end{bmatrix}, \quad 
    Z := \begin{bmatrix} 1 & 0 \\ 0 & -1 \end{bmatrix}, \quad \text{and} \quad 
    I := \begin{bmatrix} 1 & 0 \\ 0 & 1 \end{bmatrix}
\]
be the Pauli matrices (where $i=\sqrt{-1}$), and let $P := \{I, X, Y, Z\}$. The Pauli group on $n$ qubits is defined as $P_n := \{ i^e \sigma_1 \otimes \sigma_2 \otimes \cdots \otimes \sigma_n \mid \sigma_j \in P, 1\leq j\leq n, \,  e \in \{0,1,2,3\} \}$.
Up to a global phase, a single-qubit Pauli operator $\sigma \in P$ can be represented by a binary symplectic vector $[v^{(1)}, v^{(2)}] \in \mathbb{F}_2^2$ as follows:
\[
    [v^{(1)}, v^{(2)}] = \begin{cases}
        [0,0] & \text{if } \sigma = I, \\
        [1,0] & \text{if } \sigma = X, \\
        [0,1] & \text{if } \sigma = Z, \\
        [1,1] & \text{if } \sigma = Y.
    \end{cases}
\]
Since $Y=iXZ$ and $\sigma^2=I$ for all $\sigma \in P$, the multiplication of two operators in $P$ translates to the addition of their corresponding binary symplectic vectors over $\mathbb{F}_2$. Generalizing this to an $n$-qubit Pauli operator, one can represent $M = (\text{global phase})\ \sigma_1 \otimes \sigma_2 \otimes \cdots \otimes \sigma_n \in P_n$ by the binary symplectic vector 
\[
    \mathbf{m} = \big[m_1^{(1)}, m_2^{(1)}, \ldots, m_n^{(1)} \mid m_1^{(2)}, m_2^{(2)}, \ldots, m_n^{(2)} \big] \in \mathbb{F}_2^{2n},
\] 
where each $\big[ m_j^{(1)}, m_j^{(2)}\big]$ is the vector corresponding to $\sigma_j$. Thus, an $n$-qubit Pauli operator can be identified with a vector in $\mathbb{F}_2^{2n}$, while a subgroup of $P_n$ can be identified with a subspace of $\mathbb{F}_2^{2n}$.

Let $N$ be a positive integer and $\mathcal{R} := \mathbb{F}_2[z]/\langle z^N-1 \rangle$. For a polynomial $p(z) = \sum_{j=0}^{N-1} p_j z^j \in \mathcal{R}$, we denote its coefficient vector by $\pi(p(z)) := [p_0, p_1, \ldots, p_{N-1}] \in \mathbb{F}_2^N$. The Hamming weight of a polynomial is defined as $\mathrm{w}_H(p(z)) := \mathrm{w}_H(\pi(p(z)))$. The annihilator of a polynomial $p(z) \in \mathcal{R}$ is defined as $\mathrm{Ann}(p(z)) := \{r(z) \in \mathcal{R} \mid r(z)p(z) = 0 \pmod{ z^{N}-1} \} = \left\langle \frac{z^N-1}{\gcd(p(z), z^N-1)} \right\rangle$. For $p(z) \in \R$, denote its reciprocal polynomial $p^*(z) = p(z^{-1}) \pmod{z^N-1}$\footnote{There is a standard definition of reciprocal of the polynomial $p(z)$ given by $z^Dp(1/z)$, where $D$ denotes the degree of the polynomial.}. 

\subsection{Calderbank-Shor-Steane (CSS) Codes}

This manuscript focuses exclusively on quantum error-correcting codes over qubits. We assume the reader is familiar with the stabilizer formalism of quantum error-correction \cite{gottesman1997stabilizercodesquantumerror}. The Calderbank-Shor-Steane (CSS) construction \cite{css-calderbankShor,css-steane}, detailed below, provides a method to construct quantum stabilizer codes from two classical binary linear codes.

\begin{definition}[CSS Code]
    Let $n, r_1, r_2 \in \mathbb{Z}_+$, and let $\mathbf{H}_X \in \mathbb{F}_2^{r_1 \times n}$ and $\mathbf{H}_Z \in \mathbb{F}_2^{r_2 \times n}$ be matrices such that $\mathbf{H}_X\mathbf{H}_Z^T = \mathbf{0}$. Let $\mathcal{C}_X := \ker(\mathbf{H}_X)$ and $\mathcal{C}_Z := \ker(\mathbf{H}_Z)$, and let $\mathcal{C}_X^\perp$ and $\mathcal{C}_Z^\perp$ denote their respective dual codes. Define the set
    \[
        S := \left\{ [\mathbf{x} \mid \mathbf{z}] \in \mathbb{F}_2^{2n} \mid \mathbf{x} \in \mathcal{C}_X^\perp, \, \mathbf{z} \in \mathcal{C}_Z^\perp \right\}.
    \]
    The quantum code $\mathrm{CSS}(\mathbf{H}_X, \mathbf{H}_Z) \subseteq \mathbb{C}^{2^n}$ is the stabilizer code with $S$ as its stabilizer group.
\end{definition}
 Note that the condition $\mathbf{H}_X\mathbf{H}_Z^T = \mathbf{0}$ is equivalent to saying $\mathcal{C}_X^\perp \subseteq \mathcal{C}_Z$ or $\mathcal{C}_Z^\perp \subseteq \mathcal{C}_X$.
\begin{proposition}
    The dimension of $\mathrm{CSS}(\mathbf{H}_X, \mathbf{H}_Z)$ is $k = k_X + k_Z - n$, where $k_X$ and $k_Z$ are the dimensions of $\mathcal{C}_X$ and $\mathcal{C}_Z$, respectively. The minimum distance of the code is $d = \min(d_X, d_Z)$, where 
    \[
        d_X = \min_{\mathbf{c} \in \mathcal{C}_Z \setminus \mathcal{C}_X^\perp} \mathrm{w}_H(\mathbf{c}) \quad \text{and} \quad d_Z = \min_{\mathbf{c} \in \mathcal{C}_X \setminus \mathcal{C}_Z^\perp} \mathrm{w}_H(\mathbf{c})
    \]
    are the $X$-distance and $Z$-distance of the code, respectively.
\end{proposition}
We use the term dimension for the number of logical qubits encoded by a quantum code. The following proposition describes the stabilizers and the logical-$X$ and logical-$Z$ operators of the code.

\begin{proposition}
    The stabilizer group $S$ of the $\mathrm{CSS}(\mathbf{H}_X, \mathbf{H}_Z)$ code can be viewed as a direct sum of $X$- and $Z$-stabilizers, $S = S_X \oplus S_Z$, where 
    \[
        S_X := \{[\mathbf{x} \mid \mathbf{0}] \mid \mathbf{x} \in \mathcal{C}_X^\perp\} \quad \text{and} \quad S_Z := \{[\mathbf{0} \mid \mathbf{z}] \mid \mathbf{z} \in \mathcal{C}_Z^\perp\}.
    \]
    The logical-$X$ and logical-$Z$ operators of the code are given by the sets
    \[
        L_X := \{[\mathbf{z} \mid \mathbf{0}] \mid \mathbf{z} \in \mathcal{C}_Z\} \quad \text{and} \quad L_Z := \{[\mathbf{0} \mid \mathbf{x}] \mid \mathbf{x} \in \mathcal{C}_X\}.
    \]
\end{proposition}

To simplify notation in the context of CSS codes, it is convenient to represent purely $X$-type and purely $Z$-type Pauli operators using only the nonzero half of their binary symplectic vectors. For example, an $X$-type operator with the symplectic representation $[\mathbf{x} \mid \mathbf{0}]$ is denoted simply by the vector $\mathbf{x}$. An analogous convention applies to $Z$-type operators. We adopt this simplified notation throughout the remainder of this manuscript.

\begin{definition}[Quantum LDPC CSS Code]
    A family of CSS codes is called a quantum low-density parity-check (qLDPC) code family if the row and column weights of the corresponding parity-check matrices, $\mathbf{H}_X$ and $\mathbf{H}_Z$, are bounded by constants independent of the code length.
\end{definition}

Among the various qLDPC CSS code constructions known in the literature, those exhibiting good parameters in the short blocklength regime ($n \lesssim 300$ qubits), such as BB and GB codes, have drawn significant attention. In this paper, we characterize the logical Pauli operators of GB codes as modules over a polynomial ring with an explicit set of generators, derive an expression for the minimum distance, and devise a construction with bounds on the minimum distance.

\begin{definition}[Generalized Bicycle Codes]
Let $N \in \mathbb{Z}_+$ and $a(z), b(z) \in \mathcal{R}=\mathbb{F}_2[z]/\langle z^N-1\rangle$ be nonzero polynomials. Define the circulant matrices
\begin{equation*}
\mathbf{A} = \begin{bmatrix}
    \pi(a(z))\\
    \pi(za(z))\\
    \vdots \\
    \pi(z^{N-1}a(z))
\end{bmatrix} \qquad \text{ and } \qquad
\mathbf{B} = \begin{bmatrix}
    \pi(b(z))\\
    \pi(zb(z))\\
    \vdots \\
    \pi(z^{N-1}b(z))
\end{bmatrix}.
\end{equation*} 
Taking the parity-check matrices as $\mathbf{H}_X = [\mathbf{A} \mid \mathbf{B}]$ and $\mathbf{H}_Z = [\mathbf{B}^T \mid \mathbf{A}^T]$, the generalized bicycle code associated with $a(z)$ and $b(z)$ is defined as
\begin{equation*}
    \mathrm{GB}(a(z), b(z)) = \mathrm{CSS}(\mathbf{H}_X, \mathbf{H}_Z).
\end{equation*}
\end{definition}

\begin{proposition}[{\cite[Proposition~1]{Panteleev2021degeneratequantum}}]\label{prop:GBDimension}
    The code $\mathrm{GB}(a(z), b(z))$ is a $[[2N, 2\cdot \deg(h(z))]]$ quantum code, where $h(z) = \gcd(a(z), b(z), z^N-1)$.
\end{proposition}

If $h(z) = 1$, then the resulting code $\mathrm{GB}(a(z), b(z))$ is a $[[2N, 0]]$ quantum code, which encodes no logical qubits and is therefore excluded from consideration.

\begin{proposition}[\cite{GBCodeDistance_symmetry}]\label{prop:GBminDistance}
    The minimum distance of a GB code is given by
    \begin{equation*}
        d_{\min} = d_X = d_Z,
    \end{equation*}
    where $d_X$ and $d_Z$ are the $X$- and $Z$-distances of the code, respectively.
\end{proposition}
A GB code that can be decomposed as a direct sum of several GB codes of smaller blocklengths is called a commensurate GB code.

\begin{proposition}[{\cite[Statement~4]{GBCodeDistance_symmetry}}]\label{Prop:commensurate}
    A GB code $\mathrm{GB}(a(z), b(z))$ of blocklength $2N$ defined by $a(z) = a_0(z^\Delta), b(z) = b_0(z^\Delta)$, where $N = N_0\Delta$ is equivalent to the direct sum of $\Delta$ copies of the $\mathrm{GB}(a_0(z), b_0(z))$ of blocklength $2N_0$. 
\end{proposition}

An immediate consequence of Proposition~\ref{Prop:commensurate} is that, for a GB code to be incommensurate, we can assume that the $\gcd$ of $N$ and the exponents of $z$ in $a(z)$ and $b(z)$ must be $1$.
\begin{remark}
    One can construct a family of qLDPC GB codes by choosing $a(z)$ and $b(z)$ to be sparse.
\end{remark}

We partition the physical qubits into two blocks of $N$ qubits each and map each Pauli-$X$ (or Pauli-$Z$) operator to an ordered pair of polynomials in $\mathcal{R}$, representing the operator's support across the two blocks. A Pauli-$X$ operator of the form $[x_0^{(1)}, x_1^{(1)}, \ldots, x_{N-1}^{(1)}, x_0^{(2)}, x_1^{(2)}, \ldots, x_{N-1}^{(2)}]$ is identified with $(x^{(1)}(z),x^{(2)}(z)) \in \mathcal{R}^2$, where $x^{(j)}(z)=\sum_{r=0}^{N-1}{x_r^{(j)}z^r}$. This algebraic representation allows us to express the $X$-stabilizers, $Z$-stabilizers, logical-$X$ and logical-$Z$ operators as submodules of $\mathcal{R}^2$ over the ring $\mathcal{R}$. Unless stated otherwise, all operations involving polynomials from $\mathcal{R}$ are assumed to be modulo $z^N-1$.

Describing the $X$-stabilizers and $Z$-stabilizers as submodules of $\mathcal{R}^2$ is a standard result \cite[Theorem 2.1]{davenport2026generalized}. Here, we provide an independent proof for our setup.
\begin{proposition}\label{prop:stab_module}
For a GB code defined by polynomials $a(z), b(z) \in \mathcal{R}$, the $X$-stabilizers correspond to the submodule $\mathcal{M}_X := \langle (a(z), b(z)) \rangle \subseteq \mathcal{R}^2$. Similarly, the $Z$-stabilizers correspond to the submodule $\mathcal{M}_Z := \langle (b^*(z), a^*(z)) \rangle \subseteq \mathcal{R}^2$.
\end{proposition}

\begin{proof}
By definition, the $X$-stabilizers are generated by the rows of ${\mathbf{H}}_X$. Because polynomial multiplication in $\mathcal{R}$ accounts for cyclic shifts and linear combinations, any $X$-stabilizer can be written as $[\pi(u(z)a(z)) \mid \pi(u(z)b(z))]$ for some $u(z) \in \mathcal{R}$. Under our algebraic mapping, this corresponds to the element $(u(z)a(z), u(z)b(z)) \in \mathcal{R}^2$. Conversely, any element in the submodule $\langle (a(z), b(z)) \rangle$ takes the form $(v(z)a(z), v(z)b(z))$ for some $v(z)\in \mathcal{R}$ and therefore corresponds to a valid $X$-stabilizer in the rowspan of ${\mathbf{H}}_X$. Thus, the $X$-stabilizers map to the submodule $\mathcal{M}_X$. A similar argument applies to the $Z$-stabilizers as well.
\end{proof}

\section{Main Results}\label{sec:main}
We first describe the $X$-type logical operators of GB codes that commute with the
stabilizers and identify which of them act trivially on the code space.
The resulting polynomial-weight minimization will be used to prove the
distance of the construction in Section~\ref{sec:explicit}. We then
specialize to $a(z)=1+z^l$ with $l\mid N$ and $b(z)=a(z)p(z)$, which fixes the number of
encoded qubits at $2l$ while retaining freedom in the check pattern.

\subsection{Logical Pauli Operators and Minimum Distance of GB Codes}

We begin with giving a set of explicit generators for the logical-$X$ operators of a GB code, specified by the polynomials $a(z)$ and $b(z)$.
\begin{lemma}\label{lem:aDivides}
Let $C=\langle (a(z),b(z))\rangle\subseteq \mathcal{R}^2$ be a submodule. Then there exist polynomials $a'(z),b'(z)\in \mathcal{R}$ such that $a'(z)\mid z^N-1$ and  $C=\langle(a'(z),b'(z))\rangle$.  
\end{lemma}

\begin{proof}
Let $z^N-1=\prod\limits_{i=1}^r p_i(z)^{e_i}$ be the prime factorization where $p_1(z),\ldots, p_r(z)$ are irreducible factors of $z^N-1$. Define $a'(z)=\gcd(a(z),z^N-1)$ and $a_1(z)=\frac{a(z)}{a'(z)}$. We can write 
\begin{align*}  
    a'(z)&=\prod_{i=1}^rp_i(z)^{t_i},\ \ \ t_i\leq e_i,\\
    f_1(z)&=\frac{z^N-1}{a'(z)}=\prod_{i=1}^rp_i(z)^{e_i-t_i}.
\end{align*}
Since $\gcd(a_1(z),f_1(z))=1$, there exists some $u_0(z)\in \mathcal{R}$ such that $\gcd(u_0(z),f_1(z))=1$ and 
\begin{equation}\label{lem-eq2}
    a_1(z)u_0(z)= 1 \pmod{f_1(z)}.
\end{equation} 
Multiplying by $a'(z)$, we get 
\begin{align}\label{lem-eq-1}
   a(z)u_0(z)= a'(z) \pmod{z^N-1}. 
\end{align}

Define $I=\{i:\ t_i<e_i\}$ and $J=\{i:\ t_i=e_i\}$. Let 
$F_I(z)=\prod_{i\in I}p_i(z)^{e_i}$ and $F_J(z)=\prod_{i\in J}p_i(z)^{e_i}$. Then $\gcd(F_I(z),F_J(z))=1$ and $F_I(z)F_J(z)=z^N-1$. If we consider the following linear system of equations
\begin{align*}
    x= u_0(z) \pmod{F_I(z)},\\
    x=  1 \pmod{F_J(z)},
\end{align*}
 then by the Chinese Remainder Theorem (CRT), there exists $u(z)\pmod{z^N-1}$ such that $u(z)= u_0(z) \pmod{F_I(z)}$ and $ u(z)=  1 \pmod{F_J(z)}$. Consequently, $a_1(z)u(z)= a_1(z)u_0(z)=1 \pmod{f_1(z)}$, using Eq.~\eqref{lem-eq2}. On multiplying by $a'(z)$, we get  
\begin{equation}\label{lem-eq3}
    a(z)u(z)= a'(z)\pmod{z^N-1}. 
\end{equation}
Next, we claim that $u(z)$ is a unit in $\mathcal{R}$, i.e., there exists some $v(z) \in \mathcal R$ such that $v(z)u(z)= 1 \pmod{z^N-1}$. It suffices to show that no irreducible factor of $z^N-1$ divides $u(z)$. Let $p_i(z)$ be any irreducible factor of $z^N-1$. We distinguish two cases based on whether $i \in I$ or  $i \in J$. If $i\in I$, then $t_i<e_i$. Hence, the polynomial $p_i(z)$ divides $f_1(z)$, and therefore
$a_1(z)u_0(z)=1\pmod{p_i(z)}$ using Eq.~\eqref{lem-eq2}. It implies that $p_i(z)$ does not divide $u_0(z)$. Since $p_i(z)$ divides $F_I(z)$, and $u(z)= u_0(z) \pmod{F_I(z)}$, $u(z)= u_0(z)\pmod{p_i(z)}$. Consequently, $p_i(z)$ does not divide $u(z)$. On the other hand, if $i\in J$, then, $p_i(z)$ divides $F_J(z)$.
Since $u(z)= 1 \pmod{F_J(z)}$, $u(z)=1\pmod{p_i(z)}$.
Hence $p_i(z)$ does not divide $u(z)$.

Thus,  $v(z)u(z)= 1 \pmod{z^N-1}$ for some $v(z)\in \mathcal{R}$. Let $C'=\langle (a'(z),u(z)b(z))\rangle$. Then 
$(a'(z),u(z)b(z))=u(z)(a(z),b(z))$ (by Eq.~\eqref{lem-eq3}) implies that $C'\subseteq C$. Also, $(a(z),b(z))=v(z)u(z)(a(z),b(z))=v(z)(a'(z),u(z)b(z))\in C'$ implying that $C\subseteq C'$. Hence, $C=C'$. This completes the proof with $b'(z)=u(z)b(z)$. 
\end{proof}

\begin{remark}
  By Lemma~\ref{lem:aDivides},  without loss of generality, we can assume that $a(z) \mid z^N-1$ for any GB code defined by $a(z) $ and $b(z)$.

\end{remark}

Now onwards, we assume that $a(z)\mid (z^N-1)$. Moreover, from Proposition \ref{prop:GBDimension}, we assume that $\deg(h(z)) \geq 1$. Next, we describe the logical-$X$ operators as a submodule of $\mathcal{R}^2$.

\begin{proposition}\label{prop:logical_modules}
The logical-$X$ operators of the GB code defined by $a(z)$ and $b(z)$ correspond to the submodule $\mathcal{L}_X = \big\langle (a_1(z), b_1(z)), (0, c(z)) \big\rangle \subseteq \mathcal{R}^2$, where  $$a_1(z) := \frac{a(z)}{h(z)}, \quad b_1(z) := \frac{b(z)}{h(z)}, \quad c(z) := \frac{z^N-1}{a(z)}, \text{ and } h(z) := \gcd(a(z), b(z)).$$
\end{proposition}

\begin{proof}
    Under the polynomial representation, the logical-$X$ operators correspond to the dual submodule of $\mathcal{M}_Z$, that is,
    \begin{multline*}
    \mathcal{L}_X := \{ (p_1(z), p_2(z)) \in \mathcal{R}^2 \mid \pi(p_1(z)) \cdot \pi(c_1(z)) + \pi(p_2(z)) \cdot \pi(c_2(z)) = 0 \\ 
     \text{ for all } (c_1(z), c_2(z)) \in \mathcal{M}_Z \}.
    \end{multline*}
    Recall that $\mathcal{M}_Z = \langle (b^*(z), a^*(z)) \rangle = \{ (v(z)b(z^{-1}), v(z)a(z^{-1})) : v(z) \in \mathcal{R} \}$ (see Proposition~\ref{prop:stab_module}).
    Let $[p(z)]_0$ denote the constant term of a polynomial $p(z) \in \mathcal{R}$. Using the property that $\pi(p(z)) \cdot \pi(c(z)) = [p(z)c(z^{-1})]_0$, we have 
\begin{align*}
\mathcal{L}_X &= \left\{ (p_1(z), p_2(z)) \in \mathcal{R}^2 \mid [v(z^{-1})(p_1(z) b(z) + p_2(z)a(z))]_0 = 0 \text{ for all } v(z) \in \mathcal{R} \right\}.
\end{align*}
Since the inversion $v(z)\mapsto v(z^{-1})$ permutes $\mathcal{R}$, we write
\begin{align*}
   \mathcal{L}_X &=\left\{ (p_1(z), p_2(z)) \in \mathcal{R}^2 \mid [v(z)(p_1(z) b(z) + p_2(z)a(z))]_0 = 0 \text{ for all } v(z) \in \mathcal{R} \right\}\\
    & = \left \{ (p_1(z), p_2(z)) \in \mathcal{R}^2 \mid p_1(z) b(z) + p_2(z)a(z) = 0 \text{ in } \mathcal{R} \right\}.
\end{align*}
The last step follows because if $p_1(z) b(z) + p_2(z)a(z)$ has some nonzero term $z^j$, one can choose $v(z)=z^{N-j}$ to yield a nonzero constant term. 
It is easy to verify that $(a_1(z),b_1(z))$ and $(0,c(z))$ are in $\mathcal{L}_X$, so that $\mathcal{L}_X \supseteq \langle (a_1(z), b_1(z)), (0, c(z)) \rangle$. For the converse, consider any $(p_1(z), p_2(z)) \in \mathcal{L}_X$. The condition $ p_1(z) b(z) + p_2(z)a(z) = 0 $ in $\mathcal{R}$ reduces to 
 $$
 h(z)\left( p_1(z)b_1(z) + p_2(z) a_1 (z)\right) = 0 \pmod{z^N-1}, 
 $$
 which is equivalent to $p_1(z)b_1(z) + p_2(z) a_1(z) = 0 \pmod{M(z)}$, where $M(z) := (z^N-1)/h(z)$.
 Since $a_1(z) \mid M(z)$ and $\gcd(a_1(z),b_1(z)) = 1$, we obtain that $a_1(z) \mid p_1(z)$. Hence, we can write $p_1(z) = u(z)a_1(z)$ for some $u(z) \in \mathbb{F}_2[z]$. Substituting this into the above equation, we get
$$
a_1(z) \left( u(z)b_1(z) + p_2(z) \right) = 0 \pmod{M(z)}.
$$
It implies that $u(z)b_1(z) + p_2(z) = 0\pmod{M(z)/a_1(z)}$. Hence, there exists some $v(z) \in \mathbb{F}_2[z]$ such that
\begin{align*}
u(z)b_1(z) + p_2(z) = v(z)\frac{M(z)}{a_1(z)} &= v(z) c(z).
\end{align*}

Therefore,
\begin{align*}
    p_1(z) &= u(z)a_1(z),\\
    p_2(z) &= u(z)b_1(z) + v(z)c(z),
\end{align*}

and we conclude that 
\begin{equation}
    \mathcal{L}_X \subseteq \langle (a_{1}(z), b_1(z)), (0, c(z)) \rangle.
\end{equation}Hence, $\mathcal{L}_X = \left\langle (a_{1}(z), b_1(z)), (0, c(z)) \right\rangle$.
\end{proof}

One can choose another set of polynomials $\tilde{a}(z)$ and $\tilde{b}(z)$ such that $\tilde{b}(z) \mid (z^N-1)$ and $\mathrm{GB}(\tilde{a}(z), \tilde{b}(z)) = \mathrm{GB}(a(z), b(z))$~(see Lemma~\ref{lem:aDivides}). With this, a characterization of the logical-$Z$ operators can be derived analogously to Proposition~\ref{prop:logical_modules}. Since our objective is to use this characterization to analyze the minimum distance, and GB codes have the same $X$- and $Z$- distances, the description of logical-$X$ operators suffices
for this work.

Having algebraically characterized the $X$-stabilizers and logical-$X$ operators, we now evaluate the $X$-distance of the code, $d_X$. By Proposition~\ref{prop:GBminDistance}, $d_X$ dictates the overall minimum distance of the code. Since $d_X$ is the minimum weight of a logical-$X$ operator that is not an $X$-stabilizer, we must establish an algebraic condition to determine when an element of $\mathcal{L}_X$ belongs to $\mathcal{M}_X$.

\begin{lemma}\label{lem:logXmeetStabX}
        With the notations of Proposition \ref{prop:logical_modules}, let $(u(z)a_1(z), u(z)b_1(z)+v(z)c(z)) \in \mathcal{L}_{X}$ for some $u(z), v(z) \in \mathcal R$. Then,  $$(u(z)a_1(z), u(z)b_1(z)+v(z)c(z)) \in \mathcal{M}_X \iff h(z) \mid u(z) \  \text{ and } \ h(z) \mid v(z).$$
\end{lemma}

\begin{proof}
    Assume $(u(z)a_1(z), u(z)b_1(z)+v(z)c(z)) \in \mathcal{M}_X$. By definition, there exists some $u'(z) \in \mathcal{R}$ such that
    $$
    (u'(z)a(z), u'(z)b(z)) = (u(z)a_1(z), u(z)b_1(z)+v(z)c(z)).
    $$
    Comparing the first coordinates, we have $u'(z)a(z) = u(z)a_1(z)$. Since $a(z) = a_1(z)h(z)$, this yields $a_1(z)(u(z) + u'(z)h(z)) = 0 \pmod{z^N-1}$. Thus,
    $$
    u(z) + u'(z)h(z) \in \mathrm{Ann}(a_1(z)) = \langle h(z)c(z) \rangle.
    $$
    This implies $u(z) = u'(z)h(z) + t(z)h(z)c(z)$ for some $t(z) \in \mathcal{R}$, which shows $h(z) \mid u(z)$. 
    
    Now, comparing the second coordinates yields $u'(z)b(z) = u(z)b_1(z) + v(z)c(z)$. Substituting our expression for $u(z)$, we get
    $$
        t(z)b(z)c(z) + v(z)c(z)=0 \pmod{z^N-1}.
    $$
    Therefore, $t(z)b(z) + v(z) \in \mathrm{Ann}(c(z)) = \langle a(z) \rangle$. Since $a(z) = h(z)a_1(z)$, this forces $v(z)$ to be a multiple of $h(z)$.

    To see the converse, assume $h(z) \mid u(z)$ and $h(z) \mid v(z)$, so $u(z) = \tilde{u}(z)h(z)$ and $v(z) = \tilde{v}(z)h(z)$ for some $\tilde{u}(z), \tilde{v}(z) \in \mathcal{R}$. Consider the element
    $$
    \Big(u(z)a_1(z), u(z)b_1(z)+v(z)c(z)\Big)=\Big(\tilde{u}(z)a(z), \tilde{u}(z)b(z) + \tilde{v}(z)h(z)c(z)\Big) \in \mathcal{L}_X.
    $$
    Since $\gcd(a_1(z), b_1(z)) = 1$, by Bézout's identity, there exist polynomials $\hat{a}(z), \hat{b}(z) \in \mathcal{R}$ such that $\hat{a}(z)a_1(z) + \hat{b}(z)b_1(z) = 1$. Multiplying this by $\tilde{v}(z)h(z)c(z)$ gives
    $$
    \tilde{v}(z)h(z)c(z) = \hat{a}(z)\tilde{v}(z)h(z)c(z)a_1(z) + \hat{b}(z)\tilde{v}(z)h(z)c(z)b_1(z).
    $$
    Note that $h(z)a_1(z)c(z) = a(z)c(z) = 0$ in $\mathcal{R}$. Therefore, the first term vanishes, leaving $\tilde{v}(z)h(z)c(z) = \tilde{v}(z)\hat{b}(z)c(z)b(z)$. Let $u'(z) = \tilde{u}(z) + \tilde{v}(z)\hat{b}(z)c(z)$. Then
    \begin{align*}
        u'(z)a(z) &= \tilde{u}(z)a(z) + \tilde{v}(z)\hat{b}(z)c(z)a(z) = \tilde{u}(z)a(z) = u(z)a_{1}(z)\quad (\text{since } a(z)c(z) = 0), \\
        u'(z)b(z) &= \tilde{u}(z)b(z) + \tilde{v}(z)\hat{b}(z)c(z)b(z) = \tilde{u}(z)b(z) + \tilde{v}(z)h(z)c(z).
    \end{align*}
    Thus, the element $\Big(u(z)a_1(z), u(z)b_1(z)+v(z)c(z)\Big)=(u'(z)a(z), u'(z)b(z)) \in \mathcal{M}_X$.
\end{proof}

By applying Lemma~\ref{lem:logXmeetStabX} to exclude the stabilizer element from logical-$X$ operators, we obtain the following expression for the minimum distance of a GB code.

\begin{proposition}
    Let $a_1(z), b_1(z), c(z)$, and $h(z)$ be defined as in Proposition~\ref{prop:logical_modules}. The minimum distance of the GB code defined by $a(z), b(z) \in \mathcal{R}$ is given by
    \begin{equation}\label{eq:dmin_in_polynomials}
    d_{\min} = \min_{\substack{u(z),v(z) \in \mathcal{R} \\ h(z) \nmid u(z) \text{ or } h(z) \nmid v(z)}}  \mathrm{w}_H\big(u(z)a_1(z)\big) + \mathrm{w}_H\big(u(z)b_1(z) + v(z)c(z)\big).
    \end{equation}
\end{proposition}

\begin{proof}
    By Proposition~\ref{prop:GBminDistance}, the minimum distance of the code $d_{\min}=d_X$, where $d_X = \min\limits_{\mathbf{c} \in \mathcal{C}_Z\setminus \mathcal{C}_X^\perp}{w_H(\mathbf{c})}$. Writing in terms of the polynomials,
    $$d_X=\min_{(p_1(z), p_2(z))\in \mathcal{L}_X\setminus \mathcal{M}_X} \mathrm{w}_H(p_1(z))+\mathrm{w}_H(p_2(z)).$$
    Invoking Lemma~\ref{lem:logXmeetStabX} gives
\begin{equation*}
    d_X=\min_{\substack{u(z),v(z) \in \mathcal{R} \\ h(z) \nmid u(z) \text{ or } h(z) \nmid v(z)}}  \mathrm{w}_H\big(u(z)a_1(z)\big) + \mathrm{w}_H\big(u(z)b_1(z) + v(z)c(z)\big).
\end{equation*}
\end{proof}

\section{An Explicit Construction of $(2,4)$-GB Codes with Dimension and Distance $\Theta(n^{1/3})$}\label{sec:explicit}

We construct a family of $(2,4)$-GB codes by taking $a(z)=1+z^l$ and choosing $b(z)$ to be a multiple of $a(z)$ with exactly four nonzero terms. We use the distance formula obtained in the previous section to provide bounds on the minimum distance.

Let $N = lm$, and define $Z = z^l \in \R$ along with the quotient ring $\R_Z = \F[Z]/\langle Z^m - 1 \rangle$. For any polynomial $p(z) = \sum_{i=0}^{N-1} p_i z^i \in \R$, we can decompose $p(z)$ into $l$ components as follows:
\begin{align}\label{eq:l-decomp}
    p(z) &= \sum_{j=0}^{l-1} z^j \left(p_j + p_{j+l}z^l + p_{j+2l}z^{2l} + \dots + p_{j+(m-1)l}z^{(m-1)l}\right) \nonumber\\
         &= \sum_{j=0}^{l-1} z^j P^{(j)}(Z),
\end{align}
where $P^{(j)}(Z) = \sum_{i=0}^{m-1} p_{j+il} Z^i \in \R_Z$. Note that this decomposition is unique, with each component $P^{(j)}(Z) \in \R_Z$. Furthermore, because the monomials $z^j Z^i$ have disjoint supports across $j \in \{0,1,\dots,l-1\}$, the Hamming weight decomposes additively as
$$
    \wH(p(z)) = \sum_{j=0}^{l-1} \wH\big(P^{(j)}(Z)\big).
$$

\begin{lemma}\label{lem:multipleofh}
For $p(z)=\sum_{j=0}^{l-1}z^jP^{(j)}(Z)$ as in
\eqref{eq:l-decomp}, we have $(1+z^l)\mid p(z)$ if and only if
$P^{(j)}(1)=0$ for every $0\leq j\leq l-1$.
\end{lemma}
\begin{proof}
Since $z^l=1 \pmod{1+z^l}$, 
\[
p(z)=\sum_{j=0}^{l-1}z^jP^{(j)}(1)\pmod{1+z^l}.
\]
The polynomial on the right has degree less than $l$ and vanishes
exactly when all of its coefficients vanish.
\end{proof}

\begin{theorem}\label{thm:2-4}
Let $N = lm$ for some positive integers $l, m \geq 2$, and let $t = \lfloor \sqrt{m} \rfloor$. Consider
\[
    a(z) = 1 + z^l \quad \text{and} \quad b(z) = 1 + z^{\beta} + z^{tl} + z^{\beta+tl}
\]
with $1 \leq \beta \leq l-1$, and let $g = \gcd(\beta, l)$. The code $\mathrm{GB}(a(z), b(z))$ is a
\[
    \left[\!\left[2N,\, 2l,\, d \geq \min\left\{\frac{l+2\beta}{g},\, t\right\}\right]\!\right]
\]
quantum code with $(2,4)$-stabilizer weights.

Moreover, if $m = t^2$ or $\frac{l+2\beta}{g} \leq t$, then $d~=~\min\left\{\frac{l+2\beta}{g},\, t\right\}$.
\end{theorem}

\subsection{Setup and Proof Sketch for Theorem~\ref{thm:2-4}}
Since $1\le t<m$ and $1\le\beta<l$ imply $0<\beta<tl<tl+\beta<N$, the circulants $\mathbf A$ and $\mathbf B$ have row and column weights $2$ and $4$, respectively, giving every row of $\mathbf H_X$ and $\mathbf H_Z$ weight $6$ and giving the columns in the two blocks respective weights $(2,4)$ in $\mathbf H_X$ and $(4,2)$ in $\mathbf H_Z$.

For any positive integer $n$, we denote $P_n(Z) = 1 + Z + \dots + Z^{n-1} = 1 + z^l + \dots + z^{(n-1)l}$. Observe that $b(z)$ factors as
\begin{align*}
    b(z) &= 1 + z^{\beta} + z^{tl} + z^{\beta+tl} \\
         &= (1+z^l)(1+z^\beta)\big(1 + z^l + \dots + z^{(t-1)l}\big) \\
         &= (1+z^l)(1+z^\beta) P_t(Z).
\end{align*}
Since $h(z) = \gcd(a(z), b(z), z^{lm}-1) = 1 + z^l$, the code $\mathrm{GB}(a(z), b(z))$ is a $[[2N, 2l]]$ quantum code. The minimum distance of the code is given by
\begin{equation}\label{eq:construction-distance}
    d = \min_{\substack{u(z),v(z) \in \R \\ h(z) \nmid u(z) \text{ or } h(z) \nmid v(z)}} \Big(\wH\big(u(z)a_1(z)\big) + \wH\big(u(z)b_1(z) + v(z)c(z)\big)\Big),
\end{equation}
where
\[
    a_1(z) = \frac{a(z)}{h(z)} = 1, \qquad b_1(z) = \frac{b(z)}{h(z)} = (1+z^{\beta})P_t(Z), \qquad c(z) = \frac{z^N-1}{a(z)} = P_m(z^l) = P_m(Z).
\]

For any pair $u(z), v(z) \in \R$, we define $w(z) = (1+z^\beta)u(z)$ and set
\[
    x(z) = u(z)a_1(z) = u(z), \qquad y(z) = u(z)b_1(z) + v(z)c(z) = w(z)P_t(Z) + v(z)P_m(Z),
\]
with their respective $l$-component decompositions written as
\begin{align*}
    u(z) &= \sum_{j=0}^{l-1} z^j u^{(j)}(Z), & v(z) &= \sum_{j=0}^{l-1} z^j v^{(j)}(Z), & w(z) &= \sum_{j=0}^{l-1} z^j w^{(j)}(Z), \\
    x(z) &= \sum_{j=0}^{l-1} z^j x^{(j)}(Z), & y(z) &= \sum_{j=0}^{l-1} z^j y^{(j)}(Z).
\end{align*}

\textbf{Proof sketch:}
To establish Theorem~\ref{thm:2-4}, we first derive componentwise algebraic identities for $v(z)c(z)$ and $w(z)$ in Section~\ref{subsec:prelim}. We then partition all admissible pairs $(u(z), v(z))$ satisfying $h(z) \nmid u(z)$ or $h(z) \nmid v(z)$ into four exhaustive cases based on the parity evaluations $v^{(j)}(1)$ and $w^{(j)}(1)$:
\begin{enumerate}
    \setlength{\itemsep}{2pt}
    \item \textbf{Case 1 ($h(z) \nmid v(z)$, i.e., $v^{(j)}(1) = 1$ for some $j$):} Discussed in Lemma~\ref{lem:d>t}, yielding $\wH(x(z)) + \wH(y(z)) \geq t$.
    \item \textbf{Case 2 ($h(z) \mid v(z)$ and $w^{(j_0)}(1) = 1$ for some $j_0$):}  Discussed in Lemma~\ref{lem:d>t+1}, yielding $\wH(x(z)) + \wH(y(z)) \geq t+1$.
    \item \textbf{Case 3 ($h(z) \mid v(z)$, $w^{(j)}(1) = 0$ for all $j$, and $y(z) = 0$):} Discussed in Lemma~\ref{lem:y=0}, yielding $\wH(x(z)) + \wH(y(z)) \geq t$.
    \item \textbf{Case 4 ($h(z) \mid v(z)$, $w^{(j)}(1) = 0$ for all $j$, and $y(z) \neq 0$):} Discussed in Section~\ref{subsec:case4} via Proposition~\ref{prop:yneq0}, yielding $\wH(x(z)) + \wH(y(z)) \geq \min\left\{\frac{l+2\beta}{g},\, t\right\}$.
\end{enumerate}
Finally, Section~\ref{subsec:synthesis} combines these four cases with explicit upper-bound constructions to complete the proof of Theorem~\ref{thm:2-4}.

\subsection{Component Decomposition and Preliminary Identities}\label{subsec:prelim}

\begin{lemma}\label{lem:cv}
With notation as in Theorem~\ref{thm:2-4}, for any $p(Z) = \sum_{i=0}^{m-1} p_i Z^i \in \R_Z$, we have
\begin{align*}
    p(Z)P_m(Z) & = p(1)P_m(Z).
\end{align*}
In particular, we have
\[
    v(z)c(z) = \sum_{j=0}^{l-1} z^j v^{(j)}(1) P_m(Z).
\]
\end{lemma}
\begin{proof}
For any $p(Z) = \sum_{i=0}^{m-1} p_i Z^i \in \R_Z$, we have
\begin{align*}
    p(Z)P_m(Z) &= \left(\sum_{i=0}^{m-1} p_i Z^i\right)\left(\sum_{k=0}^{m-1} Z^k\right) = \sum_{i=0}^{m-1} p_i \left(\sum_{k=0}^{m-1} Z^{i+k}\right) \\
               &= \sum_{i=0}^{m-1} p_i P_m(Z) = p(1)P_m(Z) \pmod{Z^m-1}.
\end{align*}
Applying this identity to each component $v^{(j)}(Z)$ yields
\[
    v(z)c(z) = \left(\sum_{j=0}^{l-1} z^j v^{(j)}(Z)\right) P_m(Z) = \sum_{j=0}^{l-1} z^j \big(v^{(j)}(Z)P_m(Z)\big) = \sum_{j=0}^{l-1} z^j v^{(j)}(1) P_m(Z). \qedhere
\]
\end{proof}

\begin{lemma}\label{lem:w}
For any $u(z) \in \R$, the components of $w(z) = (1+z^{\beta})u(z)$ satisfy
\[
    w^{(j)}(Z) = \begin{cases}
        u^{(j)}(Z) + Z u^{(j+l-\beta)}(Z) & \text{if } 0 \leq j \leq \beta-1,\\[4pt]
        u^{(j)}(Z) + u^{(j-\beta)}(Z) & \text{if } \beta \leq j \leq l-1.
    \end{cases}
\]
In particular, evaluating at $Z=1$ gives $w^{(j)}(1) = u^{(j)}(1) + u^{((j-\beta) \pmod{l})}(1)$ for all $0 \leq j \leq l-1$.
\end{lemma}
\begin{proof}
Expanding $w(z) = (1+z^\beta)\sum_{j=0}^{l-1} z^j u^{(j)}(Z)$ and splitting the shifted sum at $j = l-\beta$, we obtain
\begin{align*}
    w(z) &= \sum_{j=0}^{l-1} z^j u^{(j)}(Z) + \sum_{j=0}^{l-1} z^{j+\beta} u^{(j)}(Z) \\
         &= \sum_{j=0}^{l-1} z^j u^{(j)}(Z) + \sum_{j=0}^{l-\beta-1} z^{j+\beta} u^{(j)}(Z) + z^l \sum_{j=l-\beta}^{l-1} z^{j+\beta-l} u^{(j)}(Z) \\
         &= \sum_{j=0}^{l-1} z^j u^{(j)}(Z) + \sum_{j=\beta}^{l-1} z^j u^{(j-\beta)}(Z) + Z \sum_{j=0}^{\beta-1} z^j u^{(j-\beta+l)}(Z) \\
         &= \sum_{j=0}^{\beta-1} z^j \big(u^{(j)}(Z) + Z u^{(j+l-\beta)}(Z)\big) + \sum_{j=\beta}^{l-1} z^j \big(u^{(j)}(Z) + u^{(j-\beta)}(Z)\big).
\end{align*}
By the uniqueness of the $l$-component decomposition~\eqref{eq:l-decomp}, the claimed expressions for $w^{(j)}(Z)$ follow immediately.
\end{proof}

\subsection{Distance Bounds for Cases 1-3}\label{subsec:cases1-3}

\begin{lemma}[Case 1]\label{lem:d>t}
Suppose $h(z) \nmid v(z)$, so that $v^{(j)}(1) = 1$ for some $0 \leq j \leq l-1$. Then
\[
    \wH(x(z)) + \wH(y(z)) \geq t.
\]
\end{lemma}

\begin{proof}
From Lemma~\ref{lem:multipleofh}, we can fix an index $j$ with $v^{(j)}(1) = 1$. By Lemmas~\ref{lem:cv} and~\ref{lem:w}, the $j$-th component of $y(z)$ is $y^{(j)}(Z) = w^{(j)}(Z)P_t(Z) + P_m(Z)$, which rearranges over $\F$ as $P_m(Z) = y^{(j)}(Z) + w^{(j)}(Z)P_t(Z)$. By the triangle inequality and submultiplicativity of Hamming weight,
\begin{align*}
    m = \wH(P_m(Z)) &\leq \wH\big(y^{(j)}(Z)\big) + \wH\big(w^{(j)}(Z)P_t(Z)\big) \nonumber\\
                    &\leq \wH\big(y^{(j)}(Z)\big) + t\,\wH\big(w^{(j)}(Z)\big) \\
                    & \leq t\left( \wH\big(y^{(j)}(Z)\big) + \wH\big(w^{(j)}(Z)\big) \right),
\end{align*}
which gives $\wH\big(y^{(j)}(Z)\big) + \wH\big(w^{(j)}(Z)\big) \geq m/t \geq t$.
Finally, by Lemma~\ref{lem:w} and the triangle inequality, $\wH\big(u^{(j)}(Z)\big) + \wH\big(u^{((j-\beta) \pmod{l})}(Z)\big) \geq \wH\big(w^{(j)}(Z)\big)$. Since $1 \leq \beta \leq l-1$, the indices $j$ and $(j-\beta) \pmod{l}$ are distinct, giving
\begin{align*}
    \wH(x(z)) + \wH(y(z)) &= \sum_{i=0}^{l-1} \wH\big(u^{(i)}(Z)\big) + \sum_{i=0}^{l-1} \wH\big(y^{(i)}(Z)\big) \\
    &\geq \wH\big(u^{(j)}(Z)\big) + \wH\big(u^{((j-\beta) \pmod{l})}(Z)\big) + \wH\big(y^{(j)}(Z)\big) \\
    &\geq \wH\big(w^{(j)}(Z)\big) + \wH\big(y^{(j)}(Z)\big) \geq t. \qedhere
\end{align*}
\end{proof}
\begin{lemma}[Case 2]\label{lem:d>t+1}
Suppose $v^{(j)}(1) = 0$ for all $j \in \{0, \ldots, l-1\}$, and there exists some $j_0 \in \{0, \ldots, l-1\}$ such that $w^{(j_0)}(1) = 1$. Then
\[
    \wH(x(z)) + \wH(y(z)) \geq t+1.
\]
\end{lemma}
\begin{proof}
Since $v^{(j)}(1) = 0$ for all $j$, Lemma~\ref{lem:cv} gives $v(z)c(z) = 0$ in $\R$, and thus $y^{(j)}(Z) = w^{(j)}(Z)P_t(Z)$ for all $j$. Write $m = qt + r$ with $0 \leq r < t$. Since $t = \lfloor \sqrt{m} \rfloor$, we have $q = \lfloor m/t \rfloor \geq t > r$.

Define $Q(Z) = 1 + Z^t + Z^{2t} + \dots + Z^{(q-1)t}$. Then $Q(Z)P_t(Z) = P_{qt}(Z)$, and we can decompose $P_m(Z) = Q(Z)P_t(Z) + R(Z)$, where $R(Z) = Z^{qt} + Z^{qt+1} + \dots + Z^{m-1}$. Observe that $\wH(Q(Z)) = q$ and $\wH(R(Z)) = m - qt = r$. Multiplying $y^{(j_0)}(Z) = w^{(j_0)}(Z)P_t(Z)$ by $Q(Z)$ gives
\begin{align*}
    y^{(j_0)}(Z)Q(Z) &= w^{(j_0)}(Z)P_t(Z)Q(Z) \\
                     &= w^{(j_0)}(Z)P_m(Z) + w^{(j_0)}(Z)R(Z) \\
                     &= P_m(Z) + w^{(j_0)}(Z)R(Z),
\end{align*}
where the last equality uses Lemma~\ref{lem:cv} and $w^{(j_0)}(1) = 1$. Rearranging gives $$P_m(Z) = y^{(j_0)}(Z)Q(Z) + w^{(j_0)}(Z)R(Z),$$
and thus 
\begin{align*}
    m = \wH(P_m(Z)) &\leq \wH\big(y^{(j_0)}(Z)\big)\wH(Q(Z)) + \wH\big(w^{(j_0)}(Z)\big)\wH(R(Z)) \\
                    &= q\,\wH\big(y^{(j_0)}(Z)\big) + r\,\wH\big(w^{(j_0)}(Z)\big).
\end{align*}
Since $w^{(j_0)}(1) = 1$, we have $\wH\big(w^{(j_0)}(Z)\big) \geq 1$. Together with $q - r \geq 1$, this yields
\begin{align*}
    \wH\big(y^{(j_0)}(Z)\big) + \wH\big(w^{(j_0)}(Z)\big) &= \frac{1}{q}\Big(q\,\wH\big(y^{(j_0)}(Z)\big) + r\,\wH\big(w^{(j_0)}(Z)\big) + (q-r)\wH\big(w^{(j_0)}(Z)\big)\Big) \\
    &\geq \frac{m + q - r}{q} = t + 1.
\end{align*}
Consequently, as in the proof of Lemma~\ref{lem:d>t},
\begin{align*}
    \wH(x(z)) + \wH(y(z)) &\geq \wH\big(u^{(j_0)}(Z)\big) + \wH\big(u^{(j_0-\beta \pmod{l})}(Z)\big) + \wH\big(y^{(j_0)}(Z)\big) \\
                          &\geq \wH\big(w^{(j_0)}(Z)\big) + \wH\big(y^{(j_0)}(Z)\big) \geq t+1. \qedhere
\end{align*}
\end{proof}

\begin{lemma}[Case 3]\label{lem:y=0}
Suppose $v^{(j)}(1) = 0 = w^{(j)}(1)$ for all $0 \leq j \leq l-1$, $(1+z^l) \nmid u(z)$, and $y(z) = 0$ in $\R$. Then
\[
    \wH(x(z)) + \wH(y(z)) \geq t.
\]
\end{lemma}
\begin{proof}
Since $v^{(j)}(1) = 0$ for all $0 \leq j \leq l-1$, we have $h(z) \mid v(z)$, and thus admissibility in~\eqref{eq:construction-distance} requires $h(z) \nmid u(z)$ (in particular, $u(z) \neq 0$). Moreover, $h(z) \mid v(z)$ implies $v(z)c(z) = 0$ in $\R$, so $y(z) = w(z)P_t(Z) = 0$. Thus $w^{(j)}(Z)P_t(Z) = 0 \pmod{Z^m-1}$ for every $0 \leq j \leq l-1$. 
We consider two cases depending on whether $w(z)$ is nonzero:
\begin{enumerate}
\item \textbf{(If $w(z) \neq 0$, i.e.,  $w^{(j_0)}(Z) \neq 0$ for some $j_0$)}: Since $w^{(j_0)}(Z)P_t(Z) = 0 \pmod{Z^m-1}$, 
\begin{equation}\label{eqn:w_j}
    w^{(j_0)}(Z)(1+Z^t) =0 \pmod{Z^m-1}.
\end{equation} Let $S = \mathrm{supp}(w^{j_0}(Z)\pmod{Z^m-1})\subseteq \{0, 1, \ldots, m-1\}$, then Eq.~\eqref{eqn:w_j} implies $S = t+S \pmod{m}$. Since $S$ is nonempty, there exists some $i\in S$ and therefore $\{i+kt \pmod{m}: k \in \mathbb{Z}\} \subseteq S$, and hence $$\wH(w^{j_0}(Z)) = |S| \geq |\{i+kt \pmod{m}: k \in Z\}| = \frac{m}{\gcd(t,m)} \geq \frac{m}{t} \geq t.$$
Therefore,
    \begin{align*}
        \wH(x(z)) + \wH(y(z)) = \wH(u(z)) &\geq \wH\big(u^{(j_0)}(Z)\big) + \wH\big(u^{(j_0-\beta \pmod{l})}(Z)\big)\\
        &\geq \wH\big(w^{(j_0)}(Z)\big) \geq t.
    \end{align*}

    \item \textbf{(If $w(z)=0$)}: We have $w(z) = (1+z^\beta)u(z) = 0$ in $\R$. Let $\delta = \gcd(\beta, N)$. Then, we have
    \[
        u(z) \in \Ann_{\R}(1+z^\beta) = \left\langle \frac{z^N-1}{\gcd(1+z^\beta, z^N-1)} \right\rangle = \left\langle \frac{z^N-1}{z^\delta-1} \right\rangle.
    \]
    Consequently, $u(z) = \alpha(z)\big(1 + z^\delta + \dots + z^{(N/\delta - 1)\delta}\big)$ for some nonzero $\alpha(z) \in \F[z]$ with $\deg \alpha(z) \leq \delta - 1$. Since $\wH(\alpha(z)) \geq 1$, disjointness of supports gives
    \[
        \wH(x(z)) + \wH(y(z)) = \wH(u(z)) \geq \frac{N}{\delta} = \frac{lm}{\delta} \geq m \geq t^2 \geq t. \qedhere
    \]
\end{enumerate}
\end{proof}

\subsection{Degree Reduction and Distance Bound for Case 4 (\texorpdfstring{$y(z) \neq 0$}{y(z) != 0})}\label{subsec:case4}

We now turn to the remaining case where $v^{(j)}(1) = 0 = w^{(j)}(1)$ for all $j$ and $y(z) = w(z)P_t(Z) \neq 0$. We first show that whenever $\wH(u(z)) < t$, a cyclic shift aligns $u(z)$ and $y(z)$ so that no wrap-around occurs modulo $z^N - 1$.

\begin{lemma}\label{lem:deg_u}
Let $N = lm$ with $l,m\geq 2$ and $t = \lfloor\sqrt{m}\rfloor$. Let $1 \leq \beta \leq l-1$ and let $0 \neq u(z) \in \R$ satisfy $\wH(u(z)) < t$. Set
\[
    y(z) = (1+z^{\beta})u(z)\left(\sum_{i=0}^{t-1} z^{il}\right) \pmod{z^N-1}.
\]
Then there exists an integer $0 \leq \gamma \leq m-1$ such that the shifted representatives $u'(z) = z^{\gamma l} u(z) \pmod{z^N-1}$ and $y'(z) = z^{\gamma l} y(z) \pmod{z^N-1}$ satisfy $\deg(u'(z)) < (m-t)l$, $\deg(y'(z)) < N$,
\[
    y'(z) = (1+z^\beta)u'(z)\left(\sum_{i=0}^{t-1} z^{il}\right) \quad \text{in } \F[z] \text{ (without modular reduction)},
\]
and $\wH(u'(z)) = \wH(u(z))$, $\wH(y'(z)) = \wH(y(z))$.
\end{lemma}
\begin{proof}
Write $u(z) = \sum_{j=0}^{l-1} z^j u^{(j)}(Z)$ and arrange the coefficients of $u(z) = \sum_{i=0}^{N-1} u_i z^i$ into an $l \times m$ array $A$ whose $j$-th row stores the coefficients of $u^{(j)}(Z)$:
\[
    A = \begin{bmatrix}
        u_0 & u_l & \dots & u_{(m-t)l} & \dots & u_{(m-1)l} \\
        u_1 & u_{l+1} & \dots & u_{(m-t)l+1} & \dots & u_{(m-1)l+1} \\
        \vdots & \vdots & \ddots & \vdots & \ddots & \vdots \\
        u_j & u_{l+j} & \dots & u_{(m-t)l+j} & \dots & u_{(m-1)l+j} \\
        \vdots & \vdots & \ddots & \vdots & \ddots & \vdots \\
        u_{l-1} & u_{2l-1} & \dots & u_{(m-t+1)l-1} & \dots & u_{ml-1}
    \end{bmatrix}.
\]
Let $s$ denote the number of nonzero columns in $A$. Since $u(z) \neq 0$, we have $1 \leq s \leq \wH(u(z)) \leq t-1$. We claim that $A$ contains at least $t$ cyclically consecutive zero columns. Suppose, to the contrary, that at most $t-1$ consecutive zero columns appear between any two adjacent nonzero columns (viewed cyclically). Then the total number of columns in $A$ is at most
\[
    m \leq s(t-1) + s = st \leq (t-1)t = t^2 - t < t^2,
\]
contradicting $t = \lfloor\sqrt{m}\rfloor$.

Since multiplying $u(z)$ by $Z = z^l$ modulo $z^N - 1$ cyclically shifts the columns of $A$ to the right by one position, there exists $0 \leq \gamma \leq m-1$ such that the array representation of $u'(z) = z^{\gamma l}u(z) \pmod{z^N-1}$ has its last $t$ columns (columns $m-t, m-t+1, \dots, m-1$) entirely zero. Thus $\deg(u'(z)) \leq (m-t)l - 1 < (m-t)l$. Furthermore, in the polynomial ring $\F[z]$, the unreduced product $(1+z^\beta)u'(z)\sum_{i=0}^{t-1} z^{il}$ has degree at most
\[
    \beta + \big((m-t)l - 1\big) + (t-1)l = ml - l + \beta - 1 < ml = N,
\]
so it equals $y'(z) = z^{\gamma l} y(z) \pmod{z^N-1}$ directly in $\F[z]$ without wrap-around. Finally, multiplication by $z^{\gamma l}$ is a coordinate permutation in $\R$, preserving both $\wH(u'(z)) = \wH(u(z))$ and $\wH(y'(z)) = \wH(y(z))$.
\end{proof}

\begin{lemma}\label{lem:y>2S}
Let $B(z) \in \F[z]$ satisfy $\deg(B(z)) < N - tl$. Define $y(z) = (1+z^{tl})B(z)$ in $\F[z]$ and $S(z) = B(z) \pmod{z^{tl}-1}$. Then
\[
    \wH(y(z)) \geq 2\,\wH(S(z)).
\]
\end{lemma}
\begin{proof}
Set $\bar{Z} = z^{tl}$ and write the $tl$-component decomposition of $B(z)$ over $\F[z]$ as
\[
    B(z) = \sum_{j=0}^{tl-1} z^j B^{(j)}(\bar{Z}).
\]
Reducing modulo $z^{tl}-1$ sets $\bar{Z} \mapsto 1$, giving
\[
    S(z) = \sum_{j=0}^{tl-1} s_j z^j \quad \text{with } s_j = B^{(j)}(1) \in \F,
\]
and $\wH(S(z)) = |\supp(S(z))|$, where $\supp(S(z)) = \{0 \leq j \leq tl-1 : B^{(j)}(1) = 1\}$. Similarly, the $tl$-component decomposition of $y(z) = (1+\bar{Z})B(z)$ in $\F[z]$ is $y(z) = \sum_{j=0}^{tl-1} z^j y^{(j)}(\bar{Z})$ with $y^{(j)}(\bar{Z}) = (1+\bar{Z})B^{(j)}(\bar{Z})$ (no wrap-around occurs since $\deg(B(z)) < N - tl$).

For any $j \in \supp(S(z))$, we have $B^{(j)}(1) = 1$, so $B^{(j)}(\bar{Z}) \neq 0$. Writing $B^{(j)}(\bar{Z}) = \sum_{i=i_{\min}}^{i_{\max}} b_i \bar{Z}^i$, where $i_{\min}$ and $i_{\max}$ are the least and greatest indices in $B^{(j)}(\bar{Z})$ such that $b_{i_{\min}}= b_{i_{\max}}=1$, the product
\[
    (1+\bar{Z})B^{(j)}(\bar{Z}) = \bar{Z}^{i_{\min}} + \sum_{i=i_{\min}+1}^{i_{\max}} (b_i + b_{i-1})\bar{Z}^i + \bar{Z}^{i_{\max}+1}
\]
contains at least the two distinct nonzero terms $\bar{Z}^{i_{\min}}$ and $\bar{Z}^{i_{\max}+1}$, so $\wH\big((1+\bar{Z})B^{(j)}(\bar{Z})\big) \geq 2$. Summing over all components yields
\begin{align*}
    \wH(y(z)) &= \sum_{j=0}^{tl-1} \wH\big((1+\bar{Z})B^{(j)}(\bar{Z})\big) \\
              &\geq \sum_{j \in \supp(S(z))} \wH\big((1+\bar{Z})B^{(j)}(\bar{Z})\big) \geq 2|\supp(S(z))| = 2\,\wH(S(z)). \qedhere
\end{align*}
\end{proof}

\begin{lemma}\label{lem:u>u_bar}
Let $u(z), B(z) \in \F[z]$ satisfy $\deg(u(z)) < N - tl$, $\deg(B(z)) < N - tl$, $(1+z^l) \nmid u(z)$, and
\[
    (1+z^{\beta})u(z) = (1+z^l)B(z) \quad \text{in } \F[z].
\]
Let $\overline{u}(z) = u(z) \pmod{z^{tl}-1}$, $S(z) = B(z) \pmod{z^{tl}-1}$, and $y(z) = (1+z^{tl})B(z)$. Then $(1+z^l) \nmid \overline{u}(z)$,
\[
    (1+z^{\beta})\overline{u}(z) = (1+z^l)S(z) \pmod{z^{tl}-1},
\]
$\wH(u(z)) \geq \wH(\overline{u}(z))$, and $\wH(y(z)) \geq 2\,\wH(S(z))$.
\end{lemma}
\begin{proof}
Since $(1+z^l) \mid (z^{tl}-1)$ and $u(z) = \overline{u}(z) + q(z)(z^{tl}-1)$ for some $q(z) \in \F[z]$, we have $(1+z^l) \nmid \overline{u}(z)$ (otherwise $(1+z^l) \mid u(z)$, a contradiction). Reducing the polynomial identity $(1+z^\beta)u(z) = (1+z^l)B(z)$ modulo $z^{tl}-1$ gives $(1+z^\beta)\overline{u}(z) =(1+z^l)S(z) \pmod{z^{tl}-1}$. Moreover, $\wH(u(z)) \geq \wH(\overline{u}(z))$ because modular reduction can only cause coefficient cancellations over $\F$, and $\wH(y(z)) \geq 2\,\wH(S(z))$ holds by Lemma~\ref{lem:y>2S}.
\end{proof}

\begin{lemma}\label{lem:u>l/g}
Under the hypotheses of Lemma~\ref{lem:u>u_bar}, we have $\wH(\overline{u}(z)) \geq l/g$. Moreover, let $T(z) = \sum_{k=0}^{l/g-1} z^{k\beta}$ and define
\[
    F(z) = S(z)T(z) \pmod{z^{tl}-1} = \sum_{j=0}^{l-1} z^j F^{(j)}(Z), \quad \text{where } F^{(j)}(Z) \in \F[Z]/\langle Z^t-1 \rangle.
\]
Then $\wH\big(F^{(j)}(Z)\big) \leq \wH(S(z))$ for all $0 \leq j \leq l-1$.
\end{lemma}
\begin{proof}
Write the $l$-component decomposition of $\overline{u}(z) \pmod{z^{tl}-1}$ as $\overline{u}(z) = \sum_{j=0}^{l-1} z^j \overline{u}^{(j)}(Z)$ with $\overline{u}^{(j)}(Z) \in \F[Z]/\langle Z^t-1 \rangle$. Since $(1+z^l) \nmid \overline{u}(z)$, there exists $0 \leq j_0 \leq l-1$ such that $\overline{u}^{(j_0)}(1) = 1$. Meanwhile, $(1+z^l) \mid (1+z^\beta)\overline{u}(z)$ modulo $z^{tl}-1$, so Lemma~\ref{lem:w} implies $\overline{u}^{(j)}(1) + \overline{u}^{(j-\beta \pmod{l})}(1) = 0$ for all $j$. Iterating this relation yields
\[
    \overline{u}^{(j)}(1) = 1 \quad \text{for all } j \in \mathcal{O} := \big\{(j_0 - k\beta) \pmod{l} : 0 \leq k \leq l/g - 1\big\}.
\]
Since $g = \gcd(\beta, l)$, the set $\mathcal{O}$ consists of $l/g$ distinct indices modulo $l$, and $\wH\big(\overline{u}^{(j)}(Z)\big) \geq 1$ for every $j \in \mathcal{O}$. Thus
\[
    \wH(\overline{u}(z)) = \sum_{j=0}^{l-1} \wH\big(\overline{u}^{(j)}(Z)\big) \geq \sum_{j \in \mathcal{O}} \wH\big(\overline{u}^{(j)}(Z)\big) \geq \frac{l}{g}.
\]

Next, write the $l$-component decomposition of $S(z)$ modulo $z^{tl}-1$ as $S(z) = \sum_{i=0}^{l-1} z^i S^{(i)}(Z)$ with $S^{(i)}(Z) \in \F[Z]/\langle Z^t-1 \rangle$, so that $\wH(S(z)) = \sum_{i=0}^{l-1} \wH\big(S^{(i)}(Z)\big)$. Fix $0 \leq j \leq l-1$. In the product $F(z) = \sum_{k=0}^{l/g-1} z^{k\beta} S(z) \pmod{z^{tl}-1}$, the $k$-th summand contributes to the $j$-th component $z^j F^{(j)}(Z)$ exclusively via the unique index $i_k = (j - k\beta) \pmod{l} \in \{0, \dots, l-1\}$. Specifically, writing $i_k + k\beta = j + q_k l$ for some integer $q_k \geq 0$, we have
\[
    F^{(j)}(Z) = \sum_{k=0}^{l/g-1} Z^{q_k} S^{(i_k)}(Z) \pmod{Z^t-1}.
\]
Since the indices $i_k = (j - k\beta) \pmod{l}$ are distinct for $0 \leq k \leq l/g - 1$, the triangle inequality and shift invariance of Hamming weight modulo $Z^t - 1$ give
\[
    \wH\big(F^{(j)}(Z)\big) \leq \sum_{k=0}^{l/g-1} \wH\big(Z^{q_k} S^{(i_k)}(Z)\big) = \sum_{k=0}^{l/g-1} \wH\big(S^{(i_k)}(Z)\big) \leq \sum_{i=0}^{l-1} \wH\big(S^{(i)}(Z)\big) = \wH(S(z)). \qedhere
\]
\end{proof}

\begin{proposition}[Case 4]\label{prop:yneq0}
Suppose $v^{(j)}(1) = 0 = w^{(j)}(1)$ for all $0 \leq j \leq l-1$, $(1+z^l) \nmid u(z)$, and $y(z) \neq 0$ in $\R$. Then
\[
    \wH(x(z)) + \wH(y(z)) \geq \min\left\{\frac{l + 2\beta}{g},\, t\right\}.
\]
\end{proposition}
\begin{proof}
Suppose for contradiction that $\wH(u(z)) + \wH(y(z)) < \min\{(l + 2\beta)/g,\, t\}$. In particular, $\wH(u(z)) < t$. Since $v^{(j)}(1) = 0$ for all $j$, we have $y(z) = (1+z^\beta)u(z)P_t(z^l) \pmod{z^N-1}$. By Lemma~\ref{lem:deg_u}, after replacing $u(z)$ and $y(z)$ by $z^{\gamma l}u(z) \pmod{z^N-1}$ and $z^{\gamma l}y(z) \pmod{z^N-1}$ (which preserves their weights and $(1+z^l) \nmid u(z)$), we may assume without loss of generality that $\deg(u(z)) < (m-t)l$ and that $y(z) = (1+z^\beta)u(z)P_t(z^l)$ holds directly in $\F[z]$ with $\deg(y(z)) < N$.

Since $w^{(j)}(1) = 0$ for all $j$, the polynomial $w(z) = (1+z^\beta)u(z)$ is divisible by $1+z^l$ in $\F[z]$ (as $\deg(w(z)) < (m-t+1)l \leq N$). Thus, we can write
\[
    (1+z^\beta)u(z) = (1+z^l)B(z) \quad \text{in } \F[z]
\]
for some $B(z)\in \mathbb{F}_2[z]$ with $\deg(B(z)) = \deg(w(z)) - l < (m-t)l = N - tl$, and consequently
\[
    y(z) = (1+z^\beta)u(z)P_t(z^l) = (1+z^l)P_t(z^l)B(z) = (1+z^{tl})B(z) \quad \text{in } \F[z].
\]
Combining Lemmas~\ref{lem:y>2S}, \ref{lem:u>u_bar}, and~\ref{lem:u>l/g}, we have for every $0 \leq j \leq l-1$:
\begin{equation}\label{eq:u_plus_y_F}
    \wH(u(z)) + \wH(y(z)) \geq \wH(\overline{u}(z)) + 2\,\wH(S(z)) \geq \frac{l}{g} + 2\,\wH\big(F^{(j)}(Z)\big).
\end{equation}

Multiplying the congruence $(1+z^\beta)\overline{u}(z) = (1+z^l)S(z) \pmod{z^{tl}-1}$ from Lemma~\ref{lem:u>u_bar} by $T(z) = \sum_{k=0}^{l/g-1} z^{k\beta}$ and using $(1+z^\beta)T(z) = 1 + z^{l\beta/g} = 1 + Z^{\beta/g}$, we obtain
\begin{align}\label{eqn:F_j}
    (1+Z) \sum_{j=0}^{l-1} z^j F^{(j)}(Z) &= (1 + Z^{\beta/g}) \overline{u}(z) \pmod{z^{tl}-1} \nonumber\\
    &= (1 + Z^{\beta/g}) \sum_{j=0}^{l-1} z^j \overline{u}^{(j)}(Z) \pmod{z^{tl}-1}.
\end{align}
By Lemma~\ref{lem:u>l/g}, there is a set $\mathcal{O}$ of $l/g$ indices $j \in \{0, \dots, l-1\}$ for which $\overline{u}^{(j)}(1) = 1$ (so $\wH(\overline{u}^{(j)}(Z))$ is an odd positive integer). If $\wH(\overline{u}^{(j)}(Z)) \geq 3$ for all $j \in \mathcal{O}$, then
\[
    \wH(u(z)) \geq \wH(\overline{u}(z)) \geq \frac{3l}{g} > \frac{l + 2\beta}{g},
\]
contradicting our initial assumption. Hence, there exists some $j_0 \in \mathcal{O}$ such that $\wH\big(\overline{u}^{(j_0)}(Z)\big) = 1$, meaning $\overline{u}^{(j_0)}(Z) = Z^h$ for some $0 \leq h < t$. Extracting the $j_0$-th component of~\eqref{eqn:F_j} in $\F[Z]/\langle Z^t-1 \rangle$ gives
\[
    (1+Z) F^{(j_0)}(Z) = (1 + Z^{\beta/g}) Z^h = (1+Z) P_{\beta/g}(Z) Z^h \pmod{Z^t-1},
\]
or equivalently,
\[
    (1+Z)\big(F^{(j_0)}(Z) + Z^h P_{\beta/g}(Z)\big) = 0 \pmod{Z^t-1}.
\]
Thus $F^{(j_0)}(Z) + Z^h P_{\beta/g}(Z) \in \Ann_{\F[Z]/\langle Z^t-1 \rangle}(1+Z) = \langle P_t(Z) \rangle = \{0, P_t(Z)\}$, which leaves two possibilities:
\[
    F^{(j_0)}(Z) = Z^h P_{\beta/g}(Z) \qquad \text{or} \qquad F^{(j_0)}(Z) = Z^h P_{\beta/g}(Z) + P_t(Z).
\]
Since $1 \leq \beta/g < l/g \leq l$, if $\beta/g < t$, then two possibilities have weights $\beta/g$ and $t - \beta/g$, respectively (and if $\beta/g \geq t$, already $\wH(u(z)) \geq l/g > \beta/g \geq t$, a contradiction). Hence
\[
    \wH\big(F^{(j_0)}(Z)\big) \in \left\{\frac{\beta}{g},\, t - \frac{\beta}{g}\right\} \implies \wH\big(F^{(j_0)}(Z)\big) \geq \min\left\{\frac{\beta}{g},\, t - \frac{\beta}{g}\right\}.
\]
Substituting this into~\eqref{eq:u_plus_y_F} for $j = j_0$ gives
\begin{equation}\label{eq:two_branch_min}
    \wH(u(z)) + \wH(y(z)) \geq \frac{l}{g} + 2\min\left\{\frac{\beta}{g},\, t - \frac{\beta}{g}\right\} = \min\left\{\frac{l + 2\beta}{g},\, \frac{l - 2\beta}{g} + 2t\right\}.
\end{equation}

From Lemma~\ref{lem:u>l/g} and $\beta < l$, we have $\beta/g<l/g\leq \wH (u(z))<t$. Therefore, $$\frac{l-2\beta}{g}+2t = t+\left (\frac{l-\beta}{g}\right )+\left (t-\frac{\beta}{g}\right )>t.$$
With this, Eq.~\eqref{eq:two_branch_min} gives $\wH(u(z))+\wH(y(z)) \geq \min\left \{\frac{l+2\beta}{g}, t\right \}$.
\end{proof}

\subsection{Upper Bound and Completion of Theorem~\ref{thm:2-4}}\label{subsec:synthesis}

Combining Cases 1-4 (Lemmas~\ref{lem:d>t}, \ref{lem:d>t+1}, \ref{lem:y=0}, and Proposition~\ref{prop:yneq0}) establishes a lower bound 
\begin{equation}\label{eqn:lowerbound}
    d \geq \min\left\{\frac{l+2\beta}{g},\, t\right\}.
\end{equation} 

\begin{remark}[Upper Bound of $(l+2\beta)/g$]\label{rem:upperbound}
For any parameters satisfying Theorem~\ref{thm:2-4}, the minimum distance always obeys
\[
    d  \leq \frac{l + 2\beta}{g}.
\]
Indeed, choose $u(z) = 1 + z^g + z^{2g} + \dots + z^{(l/g-1)g} = \frac{1+z^l}{1+z^g}$ and $v(z) = 0$. Since $g = \gcd(\beta, l) \leq \beta < l$, we have $(1+z^l) \nmid u(z)$ and $\wH(u(z)) = l/g$. Furthermore,
\[
    y(z) = \frac{1+z^l}{1+z^g}(1+z^{\beta})\big(1 + z^l + \dots + z^{(t-1)l}\big) = (1+z^{tl})\big(1 + z^g + \dots + z^{(\beta/g-1)g}\big),
\]
which has weight $\wH(y(z)) = 2(\beta/g)$, giving $\wH(u(z)) + \wH(y(z)) = \frac{l+2\beta}{g}$. 
\end{remark}

Now, we establish the two cases when the exact distance is achieved. 
\begin{itemize}
    
    \item If $\frac{l + 2\beta}{g} \le t$, then $d = \frac{l + 2\beta}{g} = \min\left\{\frac{l + 2\beta}{g}, t\right\}$ which follows from Eq.~\eqref{eqn:lowerbound} and Remark \ref{rem:upperbound}.

    \item If $m = t^2$, 
we choose $u(z) = 1 + z^{tl} + \dots + z^{(t-1)tl} = \sum_{j=0}^{t-1} Z^{jt}$ and $v(z) = 1 + z^{\beta}$. Then $(1+z^l) \nmid v(z)$, $\wH(u(z)) = t$, and in $\R$:
\begin{align*}
    y(z) &= u(z)(1+z^{\beta})P_t(Z) + (1+z^{\beta})P_m(Z) \\
         &= (1+z^{\beta}) \left[\left(\sum_{j=0}^{t-1} Z^{jt}\right)\left(\sum_{i=0}^{t-1} Z^i\right) + P_m(Z)\right] \\
         &= (1+z^{\beta}) \big[P_{t^2}(Z) + P_m(Z)\big] = 0.
\end{align*}
Thus $\wH(u(z)) + \wH(y(z)) = t$, proving $d \leq t$ and hence $d = \min\left\{\frac{l+2\beta}{g},\, t\right\}$.
\end{itemize}


\subsection{Connectedness of the Tanner graphs and indecomposability}
\label{subsec:connectivity}

We next characterize the connected components of the Tanner graphs
associated with the construction in Theorem~\ref{thm:2-4}.

\begin{proposition}[Connected components]
\label{prop:construction-connectivity}
Under the hypotheses of Theorem~\ref{thm:2-4}, with $m\geq2$, let
$g=\gcd(l,\beta)$. Each of the Tanner graphs of
$\mathbf{H}_X=[\mathbf{A}\mid\mathbf{B}]$ and
$\mathbf{H}_Z=[\mathbf{B}^{T}\mid\mathbf{A}^{T}]$ has exactly $g$
connected components. Consequently, each graph is connected if and
only if $\gcd(l,\beta)=1$.
If $g>1$, a permutation of qubits decomposes the code into a direct sum
of $g$ identical GB codes of blocklength $2N/g$, defined by
\[
    a_0(w)=1+w^{l/g},\qquad
    b_0(w)=(1+w^{\beta/g})(1+w^{tl/g})
\]
over $\mathbb{F}_2[w]/\langle w^{N/g}-1\rangle$.
\end{proposition}
\begin{proof}

Index the checks and qubits in each block by $\{0, 1, \ldots, N-1\}$. In $\mathbf{H}_X$, the two supports of row \(i\) in the left and right qubit blocks respectively are \(\{i,i+l\}\) and \(\{i,i+\beta,i+tl,i+tl+\beta\}\). A two-edge path between checks changes the index by a difference within one support, so every reachable check lies in the coset $\{i+\lambda g \pmod{N}: \lambda \in \mathbb{Z}\}$, where $g = \gcd(l,\beta)$.

Conversely, checks \(i\) and \(i+l\) share column \(i+l\) in the first block, and checks \(i\) and \(i+\beta\) share column \(i+\beta\) in the second. Traversing these paths in either direction realizes the steps \(\pm l\) and \(\pm\beta\). Thus every check in the coset is reachable. Therefore, there are $g$ connected components, each containing $N/g$ check nodes.

Negating check indices and all qubit indices modulo $N$, then swapping the two qubit blocks gives an isomorphism from this graph to the Tanner graph of $\mathbf{H}_Z$, and the same arguments follow.

To obtain the decomposition of the code, set $N_0=N/g$ and write each
index uniquely as $r+gu$, where $0\leq r<g$ and $0\leq u<N_0$.
Since $a(z)=a_0(z^g)$ and $b(z)=b_0(z^g)$, an entry of either
$\mathbf{A}$ or $\mathbf{B}$ can be nonzero only when its row and
column indices have the same residue modulo $g$. Thus the rows with
residue $r$ act only on the qubits with residue $r$ in the two qubit
blocks.

Within a fixed residue class, a shift by an exponent $gs$ acts as
\[
    r+gu\longmapsto r+g(u+s)\pmod N.
\]
Under the relabeling $r+gu\mapsto u$, this becomes the cyclic shift
$u\mapsto u+s\pmod{N_0}$. Therefore, the restrictions of
$\mathbf{A}$ and $\mathbf{B}$ to this residue class are precisely the
$N_0\times N_0$ circulants $\mathbf{A}_0$ and $\mathbf{B}_0$ defined
by $a_0(w)$ and $b_0(w)$. These restricted matrices do not depend on
$r$, which shows that the components are identical.

Order the rows of each check matrix first by $r$ and then by $u$.
For the columns, group together the qubits of residue $r$ from both
qubit blocks, retaining the order $u=0,\ldots, N_0-1$ within each
block. Use this same qubit permutation for both check matrices.
Transposition preserves the residue classes and transposes their
restricted matrices, so the reordered check matrices have the
block-diagonal forms
\[
    \mathbf{H}'_X=
      \bigoplus_{r=0}^{g-1}[\mathbf{A}_0\mid\mathbf{B}_0],
    \qquad
    \mathbf{H}'_Z=
      \bigoplus_{r=0}^{g-1}[\mathbf{B}_0^{T}\mid\mathbf{A}_0^{T}].
\]
Each pair of diagonal blocks is the defining CSS pair for
$\mathrm{GB}(a_0,b_0)$ on $2N_0=2N/g$ qubits. Since no stabilizer
generator couples qubits from different residue classes, the code
decomposes into the claimed direct sum of $g$ identical GB codes.
This is the corresponding instance of the decomposition of commensurate
GB codes~\cite[Statement~4]{GBCodeDistance_symmetry}.
\end{proof}

When $g>1$, the connected components identified in Proposition~\ref{prop:construction-connectivity} give an explicit decomposition of the code into smaller GB codes. When $g=1$, the specified Tanner graphs are connected. Since these graphs depend on the choice of stabilizer generators, their connectedness alone leaves open whether the code can decompose over disjoint sets of physical qubits. The following proposition rules out such a decomposition by showing that the full support of each $X$-check row must be placed within a single component of the partition. 
\begin{proposition}\label{prop:indecomposable}
    Under the hypotheses of Theorem 1, if \(g=\gcd(l,\beta)=1\), then the code is indecomposable over disjoint sets of physical qubits.
\end{proposition}
\begin{proof}
    For $0\le r<l$, let $C_r=\{r,r+l,\ldots,r+(m-1)l\}$ be the set of first-block qubit indices congruent to $r$ modulo $l$. By Proposition~\ref{prop:stab_module} and $b(z) = a(z)p(z)$, every $X$-stabilizer has the form $(f(z),f(z)p(z))$, where $f(z)$ is a multiple of $a(z)=1+z^l$. Lemma~\ref{lem:multipleofh} then implies that the stabilizer acts on an even number of qubits in each $C_r$.

    Suppose the code decomposes as a direct sum of codes across a partition of the physical qubits into two nonempty sets $U$ and $V$. Then the restriction of any $X$-check row to $U$ or to $V$, acting as the identity outside that set, is itself an $X$-stabilizer.
    
    Fix a row $i$. The first-block qubits in the support of row $i$ have indices $i$ and $i+l \pmod{N}$. Without loss of generality, assume that first-block qubit $i$ lies in $U$ (otherwise, swap $U$ and $V$). If first-block qubit $i+l$ lies in $V$, the restriction of row $i$ to $U$ would act on exactly one qubit in $C_{i\bmod l}$, contradicting the even-parity condition. Both first-block qubits therefore lie in $U$. The restriction to $V$ has all zeros in the first block, so its second block must be zero. Hence the entire row is supported in $U$.

    Thus each $X$-check row is supported in a single part of the partition, and rows sharing a qubit must lie in the same part. Since the Tanner graph of $H_X$ is connected by Proposition 10, all physical qubits lie in one part. This contradicts the assumption that both parts are nonempty.
\end{proof}

\subsection{Explicit One-Parameter Families}
\label{subsec:one-parameter}

Theorem~\ref{thm:2-4} gives several explicit families indexed by a single
integer. First, choose $\beta=l-1$ and $m=t^2$, with $l,t\geq2$.
Then $g=\gcd(l,\beta)=1$, and the defining polynomials are
\[
    a(z)=1+z^l,\qquad
    b(z)=(1+z^{l-1})(1+z^{tl})
\]
in $\mathbb F_2[z]/\langle z^{lt^2}-1\rangle$. By
Theorem~\ref{thm:2-4}, these codes have parameters
\begin{equation}\label{eq:connected-square-family}
    [[\,2lt^2,\,2l,\,\min\{3l-2,t\}\,]].
\end{equation}
The four exponents $0,l-1,tl,tl+l-1$ of $b(z)$ are distinct modulo $lt^2$;
therefore, the two circulant blocks have row and column weights $2$ and
$4$, respectively. Each stabilizer generator has weight $6$, and each
qubit participates in six of the specified checks in total. Both Tanner
graphs are connected by Proposition~\ref{prop:construction-connectivity}.

The following choices specialize~\eqref{eq:connected-square-family} to
one-parameter families. Throughout the table, $s\geq2$ is an integer,
$m=t^2$, and $\beta=l-1$.
\begin{table}[ht]
\centering
\caption{Explicit connected weight-$6$ GB code families.}
\label{tab:one-parameter-families}
\begin{tabular}{c c c c}
\toprule
Family & $l$ & $t$ & $[[n,k,d]]$ \\
\midrule
I & $s+1$ & $s$ & $[[\,2(s+1)s^2,\,2(s+1),\,s\,]]$ \\
II & $s$ & $s$ & $[[\,2s^3,\,2s,\,s\,]]$ \\
III & $s$ & $3s-2$ & $[[\,2s(3s-2)^2,\,2s,\,3s-2\,]]$ \\
\bottomrule
\end{tabular}
\end{table}

For each family, $n=\Theta(s^3)$, $k=\Theta(s)$, and $d=\Theta(s)$.
Consequently,
\[
    k=\Theta(n^{1/3}),\qquad d=\Theta(n^{1/3}),
    \qquad k/n=\Theta(n^{-2/3}).
\]
These are therefore qLDPC families with growing dimension and distance,
although their rate tends to zero.

For every row of Table~\ref{tab:one-parameter-families}, $d=t$, so the
parameters $[[2lt^2,2l,t]]$ agree with those of a direct sum of $l$ square
toric codes with parameters $[[2t^2,2,t]]$. This is a comparison of code
parameters; the construction above provides explicit circulant check
matrices with weight-$6$ generators and connected Tanner graphs.

For $s=2,3,4$, Family II gives $[[16,4,2]]$, $[[54,6,3]]$, and
$[[128,8,4]]$, respectively; Family III gives $[[64,4,4]]$,
$[[294,6,7]]$, and $[[800,8,10]]$. These distances follow from the
theorem, without numerical distance estimation.

\section{Concluding Remarks and Future Directions}

In this work, we investigated Generalized Bicycle (GB) codes through an algebraic framework arising from their quasi-cyclic structure. By viewing the stabilizers and logical operators as submodules over the quotient ring $R=\mathbb{F}_2[z]/\langle z^N-1\rangle$, we obtained an algebraic characterization that allows the minimum distance of GB codes to be expressed in terms of the weights of polynomials in the quotient ring. This perspective provides a useful approach to studying the structural properties and parameters of GB codes.

While numerical searches have previously provided examples 
of codes with large dimension and minimum distance, we presented an explicit construction of an infinite family of GB codes with weight-$6$ stabilizer generators. In each of the three families in Table~\ref{tab:one-parameter-families}, the number of encoded qubits and the minimum distance both grow as \(\Theta(n^{1/3})\) for codes with blocklength $n$. Each code in these families is indecomposable across partitions of the physical qubits, and both of its specified Tanner graphs are connected.
Unlike the conventional $(3,3)$ split of row weights in the two circulant blocks, our construction employs an asymmetric $(2,4)$ split of stabilizer weights. 

Several interesting questions arise from this work. When the integers $l$ and $m$ are chosen to be coprime in our construction, the GB codes obtained are equivalent to the coprime case of Bivariate Bicycle (BB) codes. Extending the algebraic framework developed here to the non-coprime case of BB codes is of interest to the authors. Another direction is to investigate whether other asymmetric $(r,s)$ splits of row weights can lead to families of GB codes with improved parameters for higher check-weights. A more detailed understanding of the polynomial weight conditions governing the logical operators may also help identify 
new infinite families of GB codes with improved dimension-distance tradeoffs.

\section*{Acknowledgement}
This work was supported by the Prime Minister's Research Fellowship (PMRF) awarded to T.\ Aswanth, and by the ANRF-NPDF grant PDF/2025/002309 awarded to G.K.\ Verma. Sakshi Dang is financially supported by the Indian Institute of Science under the Institute Postdoctoral Fellowship scheme. The work of N.\ Kashyap is supported in part by the ANRF MATRICS grant ANRF/ARGM/2025/000814/MTR. 

\section*{Statement of AI Usage}
In preparing this manuscript, the AI model ChatGPT Astra 6.1 was used to address gaps in the proof of Proposition \ref{prop:yneq0} and to explore the literature. The same model was used to edit and polish the manuscript. The remaining results and arguments were entirely done by the authors. The authors take full responsibility for the correctness of the arguments and claims in this manuscript.

\bibliographystyle{IEEEtran}
\bibliography{bibliofile}

@article{shor1995scheme,
  title={Scheme for reducing decoherence in quantum computer memory},
  author={Shor, Peter W},
  journal={Phys. Rev. A },
  volume={52},
  number={4},
  pages={R2493-R2496},
  year={1995},
  publisher={APS}
}

@article{davenport2026generalized,
  title={Generalized Bicycle Codes as Cyclic Submodules and their Automorphism Structure},
  author={Davenport, AJ and Blue, John and Chuang, Isaac},
  journal={arXiv preprint arXiv:2606.05044},
  year={2026}
}

@article{wang2026coprime,
  doi = {10.22331/q-2026-02-23-2009},
  url = {https://doi.org/10.22331/q-2026-02-23-2009},
  title = {Coprime bivariate bicycle codes and their layouts on cold atoms},
  author = {Wang, Ming and Mueller, Frank},
  journal = {{Quantum}},
  issn = {2521-327X},
  publisher = {{Verein zur F{\"{o}}rderung des Open Access Publizierens in den Quantenwissenschaften}},
  volume = {10},
  pages = {2009},
  month = feb,
  year = {2026}
}

@article{Panteleev2021degeneratequantum,
  doi = {10.22331/q-2021-11-22-585},
  url = {https://doi.org/10.22331/q-2021-11-22-585},
  title = {Degenerate quantum {LDPC} codes with good finite length performance},
  author = {Panteleev, Pavel and Kalachev, Gleb},
  journal = {{Quantum}},
  issn = {2521-327X},
  publisher = {{Verein zur F{\"{o}}rderung des Open Access Publizierens in den Quantenwissenschaften}},
  volume = {5},
  pages = {585},
  month = nov,
  year = {2021}
}

@article{Pryadko,
  title = {Quantum {K}ronecker sum-product low-density parity-check codes with finite rate},
  author = {Kovalev, Alexey A. and Pryadko, Leonid P.},
  journal = {Phys. Rev. A},
  volume = {88},
  issue = {1},
  pages = {012311},
  numpages = {13},
  year = {2013},
  month = {Jul},
  publisher = {American Physical Society},
  doi = {10.1103/PhysRevA.88.012311},
  url = {https://link.aps.org/doi/10.1103/PhysRevA.88.012311}
}

@article{bravyi2024high,
  title={High-threshold and low-overhead fault-tolerant quantum memory},
  author={Bravyi, Sergey and Cross, Andrew W. and Gambetta, Jay M. and Maslov, Dmitri and Rall, Patrick and Yoder, Theodore J.},
  journal={Nature},
  volume={627},
  number={8005},
  pages={778--782},
  year={2024},
  publisher={Nature Publishing Group UK London}
}

@article{gottesman1997stabilizercodesquantumerror,
      title={Stabilizer Codes and Quantum Error Correction}, 
      author={Daniel Gottesman},
       journal={California Institute of Technology, Ph.D. Thesis},
      year={1997},
      url={https://arxiv.org/abs/quant-ph/9705052}, 
}

@article{css-steane,
  title = {Error Correcting Codes in Quantum Theory},
  author = {Steane, A. M.},
  journal = {Phys. Rev. Lett.},
  volume = {77},
  issue = {5},
  pages = {793--797},
  numpages = {0},
  year = {1996},
  month = {Jul},
  publisher = {American Physical Society},
  doi = {10.1103/PhysRevLett.77.793},
  url = {https://link.aps.org/doi/10.1103/PhysRevLett.77.793}
}

@article{css-calderbankShor,
  title = {Good quantum error-correcting codes exist},
  author = {Calderbank, A. R. and Shor, Peter W.},
  journal = {Phys. Rev. A},
  volume = {54},
  issue = {2},
  pages = {1098--1105},
  numpages = {0},
  year = {1996},
  month = {Aug},
  publisher = {American Physical Society},
  doi = {10.1103/PhysRevA.54.1098},
  url = {https://link.aps.org/doi/10.1103/PhysRevA.54.1098}
}

@article{2bga,
  title = {Quantum two-block group algebra codes},
  author = {Lin, Hsiang-Ku and Pryadko, Leonid P.},
  journal = {Phys. Rev. A},
  volume = {109},
  issue = {2},
  pages = {022407},
  numpages = {17},
  year = {2024},
  month = {Feb},
  publisher = {American Physical Society},
  doi = {10.1103/PhysRevA.109.022407},
  url = {https://link.aps.org/doi/10.1103/PhysRevA.109.022407}
}

@Article{GBCodeDistance_symmetry,
AUTHOR = {Wang, Renyu and Pryadko, Leonid P.},
TITLE = {Distance Bounds for Generalized Bicycle Codes},
JOURNAL = {Symmetry},
VOLUME = {14},
YEAR = {2022},
NUMBER = {7},
ARTICLE-NUMBER = {1348},
URL = {https://www.mdpi.com/2073-8994/14/7/1348},
ISSN = {2073-8994},
DOI = {10.3390/sym14071348}
}

@article{breuckmann2021qldpc,
  author  = {Breuckmann, Nikolas P. and Eberhardt, Jens Niklas},
  title   = {Quantum Low-Density Parity-Check Codes},
  journal = {PRX Quantum},
  volume  = {2},
  pages   = {040101},
  year    = {2021},
  doi     = {10.1103/PRXQuantum.2.040101}
}

@inproceedings{panteleev2022asymptotically,
  author    = {Panteleev, Pavel and Kalachev, Gleb},
  title     = {Asymptotically good quantum and locally testable classical {LDPC} codes},
  booktitle = {Proc. 54th Annu. ACM SIGACT Symp. Theory Comput. (STOC)},
  pages     = {375--388},
  year      = {2022}
}

@inproceedings{leverrier2022quantum,
  author    = {Leverrier, Anthony and Z{\'e}mor, Gilles},
  title     = {Quantum {T}anner codes},
  booktitle = {Proc. IEEE 63rd Annu. Symp. Found. Comput. Sci. (FOCS)},
  pages     = {872--883},
  year      = {2022}
}

@article{viszlai2023matching,
  author       = {Viszlai, Joshua and Yang, Willers and Lin, Sophia Fuhui and Liu, Junyu and Nottingham, Natalia and Baker, Jonathan M. and Chong, Frederic T.},
  title        = {Matching generalized-bicycle codes to neutral atoms for low-overhead fault-tolerance},
  journal = {IEEE QCE2025},
  volume  = {1},
  pages   = {688--699},
  year    = {2025},
  doi     = {}
}

@article{pryadko2022qdistrnd,
   title={{QD}ist{R}nd: A {GAP} package for computing the distance of
quantum error-correcting codes},
   volume={7},
   ISSN={2475-9066},
   url={http://dx.doi.org/10.21105/joss.04120},
   DOI={10.21105/joss.04120},
   number={71},
   journal={J. Open Source Softw.},
   publisher={The Open Journal},
   author={Pryadko, Leonid P. and Shabashov, Vadim A. and Kozin, Valerii K.},
   year={2022},
   month=Mar, pages={4120} }

@misc{rowshan2026logical,
  author       = {Rowshan, Mohammad and Devitt, Simon},
  title        = {Logical operator decomposition for distance analysis of bivariate bicycle codes},
  howpublished = {arXiv:2609.39395},
  year         = {2026}
}

@article{kapshikar2023hardness,
  author  = {Kapshikar, Upendra and Kundu, Srijita},
  title   = {On the hardness of the minimum distance problem of quantum codes},
  journal = {IEEE Trans. Inf. Theory},
  volume  = {69},
  number  = {10},
  pages   = {6293--6302},
  year    = {2023}
}

@article{arnault2025variant,
  author       = {Arnault, Fran{\c{c}}ois and Gaborit, Philippe and Rozendaal, Wouter and Saussay, Nicolas and Z{\'e}mor, Gilles},
  title        = {A variant of the {B}ravyi--{T}erhal bound for arbitrary boundary conditions},
   journal = {IEEE Trans. Inf. Theory},
  volume  = {72},
  number  = {1},
  pages   = {437--446},
  year    = {2026}
}

@misc{dastbasteh2025gb,
author = {Dastbasteh, Reza and Sanz Larrarte, Olatz and Moncy, Arun John and Schnabl, Paul and Crespo, Pedro M. and Etxezarreta Martinez, Josu and Otxoa, Ruben M.},
title = {Generalized Bicycle Codes with Low Connectivity: Minimum Distance Bounds and Competitive Circuit-Level Performance},
year = {2026},
eprint = {2508.09082},
archivePrefix = {arXiv},
primaryClass = {cs.IT},
version = {2},
note = {Version 2, revised November 12, 2026},
url = {https://arxiv.org/abs/2508.09082v2}
}

@article{arnault2025classification,
  author       = {Arnault, Fran{\c{c}}ois and Gaborit, Philippe and Saussay, Nicolas},
  title        = {(2,2)-{GB} codes: Classification and comparison with weight-4 surface codes},
  journal = {IEEE Trans.  Inf. Theory},
  volume  = {72},
   number  = {7},
  pages   = {5000-5024},
  year    = {2026}
}

\appendices


\end{document}